\documentclass[manuscript]{aastex} 
\usepackage{longtable}

\shorttitle{On the Force-Freeness of the Sunspot Magnetic Fields}
\shortauthors{Tiwari, Sanjiv Kumar}

\begin{document}

\title{On the Force-Freeness of the Photospheric Sunspot Magnetic Fields as
Observed from Hinode (SOT/SP)}

\author{Sanjiv Kumar Tiwari$^{1,2}$}
 \affil{$^1$Udaipur Solar Observatory, Physical Research Laboratory, Dewali,
Bari Road,\\ Udaipur-313 001, India}
\affil{$^2$Max-Planck-Institut f\"{u}r Sonnensystemforschung, Max-Planck-Str. 2,
37191 Katlenburg-Lindau, Germany\altaffilmark{*}}
\email{tiwari@mps.mpg.de}
\altaffiltext{*}{Present Address}

\begin{abstract}
A magnetic field is force-free if there is no interaction between
the magnetic field and the plasma in the surrounding atmosphere
i.e., electric currents are aligned with the magnetic field, giving
rise to zero Lorentz force. The computation of various magnetic
parameters such as magnetic energy (using virial theorem), gradient
of twist of sunspot magnetic fields (computed from the force-free
parameter $\alpha$) and any kind of extrapolations, heavily
hinge on the force-free approximation of the photospheric sunspot magnetic
fields. Thus it is of vital importance to inspect the force-free
behaviour of the sunspot magnetic fields.
The force-freeness of the sunspot magnetic fields has been
examined earlier by some researchers ending with incoherent results.
The accurate photospheric vector field measurements with high
spatial resolution are required to inspect the force-free nature of
sunspots. For this purpose we use several vector magnetograms of high
spatial resolution obtained from the Solar Optical
Telescope/Spectro-Polarimeter (SOT/SP) aboard the Hinode. Both the
necessary and the sufficient conditions for force-freeness are examined
by checking global as well as local nature of equilibrium magnetic
forces over sunspots. We find that the sunspot magnetic fields are not
much away from the force-free configuration, although they are not
completely force-free on the photosphere. The umbral and inner
penumbral fields are more force-free than the middle and the outer
penumbral fields.
During their evolution, sunspot
magnetic fields are found to maintain their proximity to force-free
field behaviour. Although a dependence of net Lorentz force components
is seen on the evolutionary stages of the sunspots, we do not find a
systematic relationship between the nature of sunspot magnetic fields and
the associated flare activity.
Further, we examine whether the fields at the
photosphere follow linear or non-linear force free conditions.
After examining this in various complex and simple sunspots, we
conclude that, in either case, the photospheric sunspot magnetic
fields are closer to satisfy the non linear force-free field approximation.
\end{abstract}

\section{Introduction}

The concept of force-free magnetic fields first came into existence
when \cite{lust54} pointed out that the cosmic magnetic fields often
follow a condition given as,
\begin{equation}\label{fff}
  (\nabla\times {\bf B})\times {\bf B}~ =~ 0.
\end{equation}
Equation 1 represents zero Lorentz force and the
field following this equation is known as force-free
magnetic field \citep{chandra57,chandra61,parker79,low82a}.
\cite{chandra56b,chandra56a}
explained many characteristics of the force-free field equation
including its solutions in axisymmetric as well as in non-axially
symmetric cases for $\alpha$ = constant. Many of the characteristics
of force-free behaviour in the solar atmosphere were discussed by
\cite{wolt58,molo69,molo74,low73a,low74a,low77a,
low78,low80,low75,low07,parker89,parker90} etc.

We can rewrite Equation 1 as,
\begin{equation}\label{fff1}
    \nabla\times {\bf B}~=~ \alpha {\bf B}
\end{equation}
which denotes that the electric current density {\bf J} ($=1/4\pi~\nabla\times{\bf B}$)
is proportional to the magnetic field {\bf B}. Here $\alpha$ is a constant
and some function of position. If we take divergence of the Equation
\ref{fff1}, as a result of Maxwell's equation
\begin{equation}\label{maxw}
\nabla\cdot {\bf B} = 0
\end{equation}
we get the following
\begin{equation}\label{fff2}
   {\bf B}\cdot\nabla\alpha =~0.
\end{equation}
Equation 4 shows that the force-free parameter $\alpha$ remains constant
along any field line. However, the $\alpha$ can vary across the field
lines. The z component of Equation \ref{fff1} allows us to compute
the distribution of $\alpha$ on the photosphere (z=0):
\begin{equation}\label{alpha}
    \alpha=[\frac{\partial B_y}{\partial x}-\frac{\partial B_x}{\partial y}]/B_z
\end{equation}
Three cases may arise: (i) $\alpha=0$ everywhere, i.e., there is no electric
current in the atmosphere resulting in a potential field (for
details please see, \cite{schm64,semel67,saku82,saku89,regn07})
(ii) $\alpha=constant$ everywhere, i.e., linear force-free field state
\citep[see e.g.,][etc]{naka72,gary89,saku89,van10} which is not always
valid and (iii) $\alpha$ varies spatially, i.e., nonlinear force-free
magnetic field
\citep[see e.g.,][etc]{saku79,low82a,amar97,amar99,amar06,wieg04,regn07,schr06,
schr08,derosa09,metc08,mack06a,mack06b,mack09,wheat09}.
This is the most common state expected \citep{low85}. However, the
vector magnetograms with high spatial resolution are required to confirm the
non-linear force-free state of sunspot magnetic fields.

The potential field approximation of sunspot fields does not appear
valid due to the presence of local vertical current distributed over
the sunspots \citep[see, e.g.][and references therein]{su09,tiw09b}. The linear
force-free field approximation has been found closely valid in a sunspot
NOAA AR 5747 studied by \cite{moon02} using relatively lesser resolution
data. In earlier works, perhaps the insufficient resolution of the data
obscured the conclusions about the validity of linear/non-linear force-free
field approximations. In the present work, we also inspect the validity of
linear or nonlinear assumptions after examining the
force-free nature of sunspot magnetic fields by using high spatial
resolution photospheric vector magnetograms obtained from Solar
Optical Telescope/Spectro-Polarimeter onboard the Hinode.

The general belief is that the photospheric field is not force-free
due to high plasma $\beta$ (ratio of plasma pressure to magnetic
pressure) and is force-free in the corona due to low plasma $\beta$
\citep[see e.g.,][]{gary01}.
However this has become possible to verify on the photosphere with
the advent of the vector magnetograms with high spatial resolution.
Observations of vector magnetograms in the chromosphere will also make it
feasible. However, such data are yet to obtain. Spatial variations of
plasma and magnetic pressures over a sunspot observed in infrared lines
have been studied by \cite{sola93a} and \cite{shibu04}. The fields of the umbra
were found more force-free than that of the penumbra.
Here, we make an attempt to find the proximity of the photospheric field
to force-free nature by using the high resolution vector magnetograms
obtained from Hinode (SOT/SP).
\cite{metc95} and \cite{moon02} have already performed some analysis
in this regard using ground based observations.
By using the observations of Stokes parameters at
six wavelengths within the spectral line of Na I $\lambda5896$,
\cite{metc95} concluded that the magnetic field for an active region
(AR) NOAA 7216 is not force-free in the photosphere whereas it
becomes force-free approximately 400 km above the photosphere in the
chromosphere. Whereas \cite{moon02}, by examining the observations
of Fe I doublet $\lambda6301.5$ and $\lambda6302.5$ for 12 vector
magnetograms of three active regions, found that the photospheric
magnetic fields are not so far from the force-free nature as
conventionally regarded. Another rigorous study in this regard was
performed by \cite{geor04}. With some limitations
of their techniques and quality of data, \cite{geor04} concluded that
the photospheric active region magnetic fields are not force-free.

All the above results are not well in agreement. One of the reasons may
be the inadequate resolution of their data sets. The accurate photospheric
vector magnetic field measurements with high spatial resolution are needed to
fully examine (both necessary and sufficient conditions) the force-free nature
of sunspots. We have used 19 high resolution vector magnetograms obtained from
the Solar Optical Telescope/Spectro-Polarimeter (SOT/SP) aboard the Hinode to
inspect the force-free behaviour of the photospheric sunspot magnetic fields.
The effect of polarimetric noise present in the data obtained from
SOT/SP does not affect much in derivation of the magnetic field
parameters \citep{tiw09a,gosain10}.

We have further analyzed total 60 vector magnetograms of four active
regions: two complex and flare-productive, namely NOAA ARs 10930 and 10960,
and two simple and relatively quiet, namely NOAA ARs 10933 and 10961.
Total of 18 vector magnetograms for the AR 10930, 12 for the AR 10933, 15 for
the ARs 10960 and 10961 each, have been analyzed to look at the
variations in the force-free behavior during their temporal evolution on
the one hand and to find any relationship with the flare activity on
the other hand. We also investigate whether linear or non-linear force-free
field approximation is closer to validity in the sunspot magnetic fields
as observed from the SOT/SP.

The paper has been organized in the following way. The methods to
test the local and global force-freeness of the photospheric sunspot
magnetic fields has been illustrated in the following section
(Section 2). In Section 3, we describe the data sets used and
methods of its analysis. Section 4 portrays the details of the results
obtained. Finally in Section 5, we present our conclusions with appropriate
discussion.

\section{Conditions for the Force-freeness of the Magnetic Fields}

Before considering the parameters which depend on the force-free
parameter $\alpha$ and perform the modeling that are
based on the force-free assumption, as discussed earlier, it is of vital
importance to examine whether the photospheric magnetic fields are
really force-free, and such approximations are realistic.
For this purpose we need to examine both global and local nature
of Lorentz forces, i.e., both the necessary and the sufficient
conditions.

\subsection{Necessary condition}

We use the following well known property of the Lorentz
force \citep{molo69,molo74,parker79,low85} that the
Lorentz force can be written as the divergence of the Maxwell stress
tensor:
\begin{equation}\label{tensor}
    M_{ij} = -\frac{B^2}{8\pi}\delta_{ij} + B_i B_j/4\pi
\end{equation}
We can convert the total force over a volume into the total Maxwell
stress exerted upon the boundary surface
\citep[e.g.,][]{parker79,low85}. Under the assumption that the
magnetic field above the plane z=0 (photosphere) falls off enough as
z goes to infinity, the net Lorentz force in the half-space z$>$0 is
just the Maxwell stress integrated over the plane z=0
\citep{aly84,low85}. Thus the components of the net Lorentz force at
the plane z=0 can be expressed by the surface integrals as follows:
\begin{eqnarray}\label{fxfyfz}
  F_x &=& -\frac{1}{4\pi}\int B_x B_z dx dy \\
  F_y &=& -\frac{1}{4\pi}\int B_y B_z dx dy \label{fy} \\
  F_z &=& -\frac{1}{8\pi}\int (B_z^2-B_x^2-B_y^2) dx dy \label{fz}
\end{eqnarray}
where $F_x$, $F_y$ and $F_z$ are the components of the
net Lorentz force. According to \cite{low85} the necessary conditions
for any magnetic field to be force-free are that
\begin{equation}\label{conditions}
\mid F_x\mid \ll F_p,~~
\mid F_y\mid \ll F_p,~~and~~
\mid F_z\mid \ll F_p.
\end{equation}
Where $F_p$ is the
force due to the distribution of the magnetic pressure on z=0, as
given by,
\begin{equation}\label{fp}
    F_p = \frac{1}{8\pi}\int (B_x^2+B_y^2+B_z^2) dx dy.
\end{equation}

Thus the inequalities given by Equation \ref{conditions} provide a test
for whether a force-free field exists at the boundary plane z = 0 (photosphere).
The criterion for the fields to be force-free is that the Equations 7, 8
and 9 normalized with respect to $F_p$ should each be much less than
unity \citep{low85}. It was discussed by \cite{metc95} that the
magnetic field is completely force-free if the aforementioned ratios are
less or equal to 0.1.

It is to be noted, however, that the inequalities given by Equation
\ref{conditions} are only necessary condition for the fields to be force-free.
The reason for this is that some information is lost in the surface
integration in Equations 7, 8 and 9.

\subsection{Sufficient condition}

We can infer the sufficient condition for the force-freeness of sunspot
magnetic fields from the distributions of vertical tension force over
sunspots \citep{venk10}. In a force-free case the tension force will
balance the gradient of magnetic pressure demanding for zero Lorentz force.
The usefulness of the tension force has not found much attention
earlier in the literature but for few studies
\citep{venk90a,venk90,venk93,venk10,tiw10}. Recently \cite{venk10} pointed
out the utility of tension force as a diagnostic of dynamical
equilibrium of sunspots.

Let us begin with the equation of magneto-hydrostatic equilibrium
\begin{equation}\label{mhse}
    (\nabla \times {\bf B}) \times {\bf B}/4\pi - \nabla p + \rho g = 0
\end{equation}
where first term is the Lorentz force, second term is the gradient of gas
pressure and the last term is the gravitational force onto the plasma.
Lorentz force (let us call it as {\bf F}) can be split into two terms as,
\begin{equation}\label{}
    {\bf F} = \frac{\bf (B \cdot \nabla) B}{4\pi} - \frac{\bf \nabla (B \cdot B)}{8\pi}
\end{equation}
The first term in the right hand side in the above equation is the
tension force (let us call it as {\bf T}). The second term represents the gradient of
the magnetic pressure i.e., the force due to magnetic pressure.
We can expand the vertical component of the tension force in terms
of the derivatives of the transverse components of the magnetic fields as
follows:
\begin{equation}\label{tz}
{T_z} = \frac{1}{4\pi}[B_x \frac{\partial B_z}{\partial x} + {B_y} \frac{\partial B_z}{\partial y} -
B_z (\frac{\partial B_x}{\partial x} + \frac{\partial B_y}{\partial y})]
\end{equation}
where, the last component in the right hand side has been drawn from the Equation \ref{maxw}.

In a force-free condition \citep{chandra61,parker79}, the tension force balances
the magnetic pressure gradient. In that condition, the Equation of magneto-hydrostatic
equilibrium reduces to hydrostatic equilibrium
\begin{equation}\label{hydsteq}
     \nabla p = \rho g
\end{equation}
which represents a state of lowest order pressure and density. In the cases
when {\bf B} is strong, the departures from the lowest order hydrostatic
pressure and density will be large. Thus a small variations in {\bf B} can lead to
large variations in plasma parameters, and could be one explanation
for existence of the variety of fine structures
\citep{su09,tiw09b,tiw09e} in the strong magnetic field system
of sunspots as observed from the high resolution data of SOT/SP.

In the Equation \ref{hydsteq}, the plasma pressure scale height will be independent
of the scale height of the magnetic field. Whereas, in a tension-free field,
the z-component of the magneto-hydrostatic equilibrium requires,
\begin{equation}\label{}
    \partial / \partial z (B^2 / 8\pi + p) = \rho g .
\end{equation}
The field in this condition cannot be force-free. In the case when
tension is zero, the scale height of the magnetic field becomes very
large, and becomes vulnerable to magnetic reconnection with the
over-lying pre-existing field system. This can lead to flare
initiation \citep{venk90a,venk90,venk93}.

We have computed tension force using Equation \ref{tz} and expressed
it in the units of dynes/cm$^3$. It was found
in the analysis of two active regions namely NOAA ARs 10933 and
10930 by \cite{venk10} that the magnitude of vertical tension force
attains values comparable to the force of gravity at several places
over the sunspots. This must mean that the non-magnetic forces will
not be able to balance this tension force. Only gradient of the
magnetic pressure can match this force. These sunspot features where
the magnetic tension balances the magnetic pressure gradient result
into the force-free configurations. This serves as sufficient condition
for verifying the force-freeness of the sunspot magnetic fields.
We have evaluated the vertical tension force over all the sunspots that we
have studied as listed in Table 1.

\section{Data Sets Used and Its Analysis}

We have used the high spatial resolution vector magnetograms of 19 active
regions (ARs) obtained from the Solar Optical
Telescope/Spectro-polarimeter (SOT/SP:
\cite{tsun08,suem08,ichi08,shim08}) onboard the Hinode \citep{kosu07}. A
series of vector magnetograms of active regions (ARs) NOAA 10930, 10933,
10960 and 10961 have also been used to investigate the temporal
changes in the force-free nature of the sunspot magnetic fields
 and also to find any relationship with the associated flare activities.

The calibration of the Hinode (SOT/SP) data have been done by using
the standard ``SP\_PREP'' routine developed by B. Lites which is
available in the Solar Software package. The ``SP\_PREP'' routine
first computes the thermal shifts in the spectral and slit dimensions,
and then applies the drift corrections for calibrating the data
from level0 to level1. The prepared polarization spectra have then
been inverted to obtain vector magnetic field components using an
Unno-Rachkowsky \citep{unno56,rach67} inversion under the assumption
of Milne-Eddington (ME) atmosphere \citep{lando82,skum87}. The ME
code used for inversion is the "STOKESFIT", which have been
developed by T. R. Metcalf and has been kindly made available in the
Solar-Software package. We have used the latest version of the inversion
code which returns the true field strengths along with the filling
factor. Some of the data sets are also used from the
Community Spectro-polarimetric Analysis
Center (CSAC; \url{http://www.csac.hao.ucar.edu/}).

Due to insensitivity of Zeeman effect to orientation of the transverse
fields, an inherent 180$^\circ$ ambiguity in the azimuth
determination will be present. Numerous techniques have been
developed and applied to resolve this problem, but still complete
resolution is not possible. The chromospheric and coronal structures
are also proposed to complement the other methods. In the data sets
which we have studied, the 180$^\circ$ azimuthal ambiguity have been removed
by using acute angle method \citep{harv69,saku85,cupe92}. Most of the data
sets used, have high spatial sampling with $\sim0.3$ arcsec/pixel. A
few samples are also observed in "Normal mode" of SOT with a
spatial sampling of $\sim0.16$ arcsec/pixel. To minimize the projection
effects, the vector fields are transformed to disk center \citep{venk89}
whenever they are much away from disk center, as is indicated in
Tables 1 \& 2.

We minimize the noise present in the data sets in the following way:
the pixels with transverse $(B_t)$ and longitudinal magnetic field
$(B_z)$ values greater than a certain level are only analysed. A
quiet Sun region is selected for each sunspot to decide this
critical threshold. Then 1$\sigma$ deviation in the three vector
field components $B_x$, $B_y$ and $B_z$ are evaluated separately.
The resultant deviation in $B_x$ and $B_y$ is taken as 1$\sigma$
noise level for transverse field components. Those pixels with
longitudinal and transverse fields simultaneously greater than twice
the above mentioned noise levels are only analysed. This method to
minimize the noise level has been successfully practiced
\cite[see, e.g.][etc.]{tiw09b,tiw09e,tiw10a,venk09,venk10,gosain09,gosain10}.

In most of the active regions studied, the flux is balanced with
less than or equal to 10\% uncertainties. But some sunspots are single
polarities with much larger imbalance of flux. Inclusion of such sunspots
are justified as follows: we compare the values of
$F_x/F_p$, $F_y/F_p$ and $F_z/F_p$ by taking whole active region
and each polarity separately for those cases when both polarities are
available. It is found that when the flux is balanced i.e.,
both polarities are taken together, the values of
$F_x/F_p$, $F_y/F_p$ and $F_z/F_p$ are smaller than the values computed
by taking the positive and negative polarities separately.
Thus, in the cases when flux is highly imbalanced, we consider
that the lower limits of the components of the net Lorentz forces are
obtained. It means they are more force-free
than they appear from the calculated values.
Further, the flux balance is not a concern for sufficient condition
as we compute the vertical tension force locally.

\section{Results}

We have studied the characteristics of Lorentz forces in the
photospheric magnetic fields of 19 sunspots obtained from the high
spatial resolution spectro-polarimetric observations by the instrument
SOT/SP aboard the {\it Hinode} satellite. The values of Lorentz force
components normalized with the respective magnetic pressure forces
are given in Table 1. It can be noticed that the values of
$|F_x/F_p|$, $|F_y/F_p|$ and $|F_z/F_p|$ in most of the cases are
approximately equal to 0.1 or, sometimes even smaller. In some cases
when they are greater than 0.1, they are always lesser than 0.6.
If the ratios $|F_x/F_p|$, $|F_y/F_p|$ and $|F_z/F_p|$ are less than unity
then the field is considered to be force-free \citep{low85,metc95,moon02}.
Thus, we can conclude that the sunspot magnetic fields are not
far from the force-free nature as has been suspected for
long time. In contrast they show much closeness to the approximation
of the force-free atmosphere in the photosphere. Our results are
consistent with that of \cite{moon02}, however, the quality of the data
used in our analysis is far better.

We can notice that $F_z/F_p$ is negative for all the sunspots, except for one.
The reason for negative $F_z/F_p$ can be understood in the following way:
The plasma pressure is weaker in magnetic field concentrations such as in the
sunspots as compared to the quiet Sun. Thus the gravity alone cannot cope with
the upward pressure gradient force over sunspots. Therefore, to maintain
the magnetohydrostatic equilibrium, downward Lorentz force $(F_z)$ is needed
to balance excessive upward pressure gradient force. However $F_z$ can be
positive or zero if the flux tubes lie very low in the atmosphere.
There is yet another plausible explanation for the negative $F_z/F_p$.
The sunspot magnetic field is confined by strong exterior photospheric
pressure. Since the sunspot field expands outward with the height, the Wilson
effect would be equivalent to the expanded field pushing the photospheric plasma
downward around the edge of the sunspot. If this effect dominates, then $F_z/F_p$
will be negative in the layer of the atmosphere where the polarimetric signals
originate. On the other hand, if the field is loaded with the mass in the
penumbral region, its weight can push the field downward giving rise to a
positive $F_z/F_p$ to support that weight. If this depressed part of the field
is within the layer of polarimetric measurement, we observe positive $F_z/F_p$
(B. C. Low 2011, private communication).

As discussed earlier, the above results are drawn merely from the
necessary conditions as these reflect only global property
of the sunspot magnetic fields. However, as sufficient condition for
the force-freeness, we need to look at the distribution of
Lorentz forces in the fine structures of sunspots.
We computed the vertical component of the magnetic tension force,
a component of the Lorentz force, by using Equation \ref{tz}.
For example, the vertical tension force distribution for two sunspots,
one quiet and one complex, are shown in the
Figures 1 and 2 respectively. The associated histograms are plotted in
Figures 1(c) and 2(c) respectively. We distinguish between the nature of
the vertical tension forces within the umbra and the penumbra, and this has
been described later in Section 4.1.

\begin{table}
\caption{\tiny List of the Active Regions, that we have studied, with their
observational details and results obtained.}
\small{} \centering
\begin{tiny}
\begin{tabular}{c c c c c c c c c}
\hline     
\hline
NOAA AR & Date: Time(UT)  & Position  &$F_x/F_p$ &$F_y/F_p$ & $F_z/F_p$& $<\alpha>\pm\mu_{\alpha}$& $<T_z>\pm\mu_{T_z}$     & comment \\
Number  & of Observation  &           &          &          &          & $\times10^{-8}m^{-1}$       & $\times10^{-3}dyn~cm^{-3}$ &         \\
\hline
10972   & 07 Oct 2007: 0200  & S05W10   &  -0.137  &   0.203  & -0.500 & $ 0.225 \pm 0.819$     & $-5.529 \pm 0.057$      & quiet,diffused   \\
10971   & 29 Sep 2007: 1011  & N03W04   &   0.006  &   0.102  & -0.508 & $ 0.648 \pm 0.758$     & $-5.038 \pm 0.053$      & quiet,diffused   \\
10970   & 05 Sep 2007: 0203  & S07W46(t)&  -0.155  &   0.213  & -0.148 & $ 2.655 \pm 0.731$     & $-3.446 \pm 0.047$      & quiet,diffused   \\
10969   & 29 Aug 2007: 0001  & S05W33(t)&   0.384  &   0.131  & -0.253 & $ 2.930 \pm 0.698$     & $-5.735 \pm 0.049$      & quiet \\
10966   & 07 Aug 2007: 0500  & S06E20(t)&  -0.028  &   0.196  & -0.345 & $-2.249 \pm 1.354$     & $-5.303 \pm 0.086$      & complex, weak  \\
10963   & 12 Jul 2007: 0001  & S06E20(t)&  -0.003  &   0.119  & -0.206 & $-3.252 \pm 0.377$     & $-4.730 \pm 0.031$      & active   \\
10961   & 01 Jul 2007: 0304  & S13E05(t)&   0.024  &   0.172  & -0.271 & $-4.754 \pm 0.939$     & $-5.753 \pm 0.072$      & quiet    \\
10960   & 07 Jun 2007: 0304  & S07E07   &   0.137  &   0.093  & -0.482 & $-1.588 \pm 0.446$     & $-5.625 \pm 0.045$      & complex, active   \\
10956   & 18 May 2007: 0809  & N03W01   &   0.248  &  -0.037  & -0.237 & $ 8.523 \pm 0.636$     & $-6.974 \pm 0.060$      & complex, rotating  \\
10955   & 13 May 2007: 1100  & S09W28(t)&   0.254  &  -0.035  & -0.349 & $ 0.411 \pm 0.689$     & $-5.143 \pm 0.063$      & quiet   \\
10953   & 29 Apr 2007: 0001  & S10E22(t)&   0.046  &   0.067  & -0.089 & $-2.499 \pm 0.609$     & $-4.424 \pm 0.058$      & quiet, strong   \\
10944   & 03 Mar 2007: 0001  & S05W30(t)&   0.078  &  -0.032  & -0.133 & $-1.749 \pm 0.284$     & $-3.731 \pm 0.029$      & quiet, strong   \\
10940   & 01 Feb 2007: 0708  & S04W05   &   0.055  &  -0.033  & -0.279 & $-1.265 \pm 0.279$     & $-7.371 \pm 0.032$      & complex   \\
10939   & 22 Jan 2007: 1314  & S04W45(t)&   0.148  &  -0.034  & -0.085 & $-1.317 \pm 0.650$     & $-5.512 \pm 0.057$      & complex     \\
10933   & 05 Jan 2007: 1213  & S04E03   &   0.045  &   0.002  & -0.108 & $-0.557 \pm 0.123$     & $-6.341 \pm 0.026$      & quiet   \\
10930   & 11 Dec 2006: 1314  & S05W07   &   0.039  &  0.038   & -0.062 & $-8.604 \pm 0.316$     & $-5.749 \pm 0.028$      & complex, rotating    \\
10926   & 03 Dec 2006: 0607  & S09W26(t)&   0.139  &   0.114  & -0.183 & $ 1.674 \pm 0.849$     & $-5.479 \pm 0.068$      & complex   \\
10923   & 10 Nov 2006: 1617  & S06E40(t)&   0.152  &   0.007  &  0.037 & $-4.840 \pm 0.761$     & $-4.903 \pm 0.059$      & complex, strong   \\
10921   & 06 Nov 2006: 1415  & S08W35(t)&   0.195  &   0.121  & -0.222 & $-1.169 \pm 0.453$     & $-4.603 \pm 0.041$      & complex    \\
\hline
(t) :  \it transformed \\
\end{tabular}
\end{tiny}
\end{table}

\subsection{Nature of magnetic field in umbra and penumbra}

It can be noticed that the magnitude of
vertical tension at many places (almost everywhere in umbra and inner
penumbra) reaches up to $10^{-2}$ dyn/cm$^3$. This value of the tension force is
comparable with the gravitational forces experienced over sunspots which
ranges from $5 \times 10^{-3}$ dyn/cm$^3$ to $10^{-2}$ dyn/cm$^3$ for quiet sun
density to sunspot umbral densities.
Such high values of tension force can not be balanced only by non-magnetic
forces i.e., gas pressure force and gravitational force.
For example, when one component of the lorentz force (say tension force)
becomes comparable to gravity, and other component (magnetic pressure force)
is not matching, then the gas pressure force needs to become double to balance both
gravity and magnetic tension. This being unlikely, we must only
rely on magnetic pressure gradient to balance it. For an estimation of
magnetic pressure gradient, let us make use of typical field strength values of
$\sim2000$ G and $\sim1000$ G for spines and intraspines
respectively in penumbra \citep[e.g.][]{borr08}.
Considering a vertical scale height $\Lambda$ of $\sim 100$ km for magnetic field
variation, we can estimate the magnetic pressure gradient ($\Delta B^2 /8 \pi
\Lambda$) which comes out to be $\sim 1.2 \times 10^{-2}$ dyn/cm$^3$.
The magnetic tension force achieved in our study is comparable to this value
at most of the places over sunspots.
Thus, at those places where tension force is high, it is balanced by gradient
of magnetic pressure leading to a force-free configuration. However, one should
remember that the magnetic fields do not show a force-free behaviour over
whole sunspot.

\begin{figure}
\epsscale{0.96}
\includegraphics[width=0.46\textwidth]{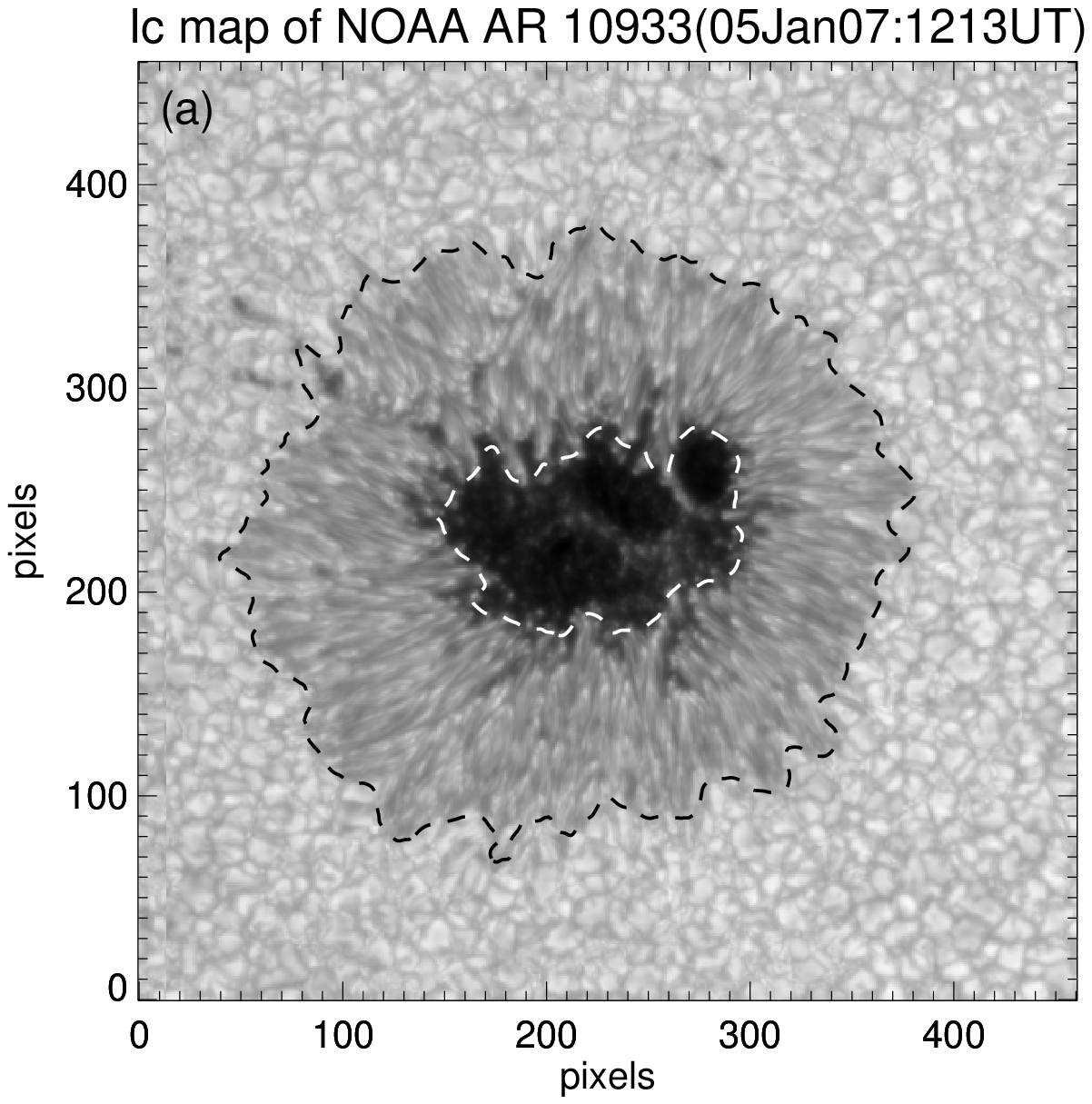}
\includegraphics[width=0.526\textwidth]{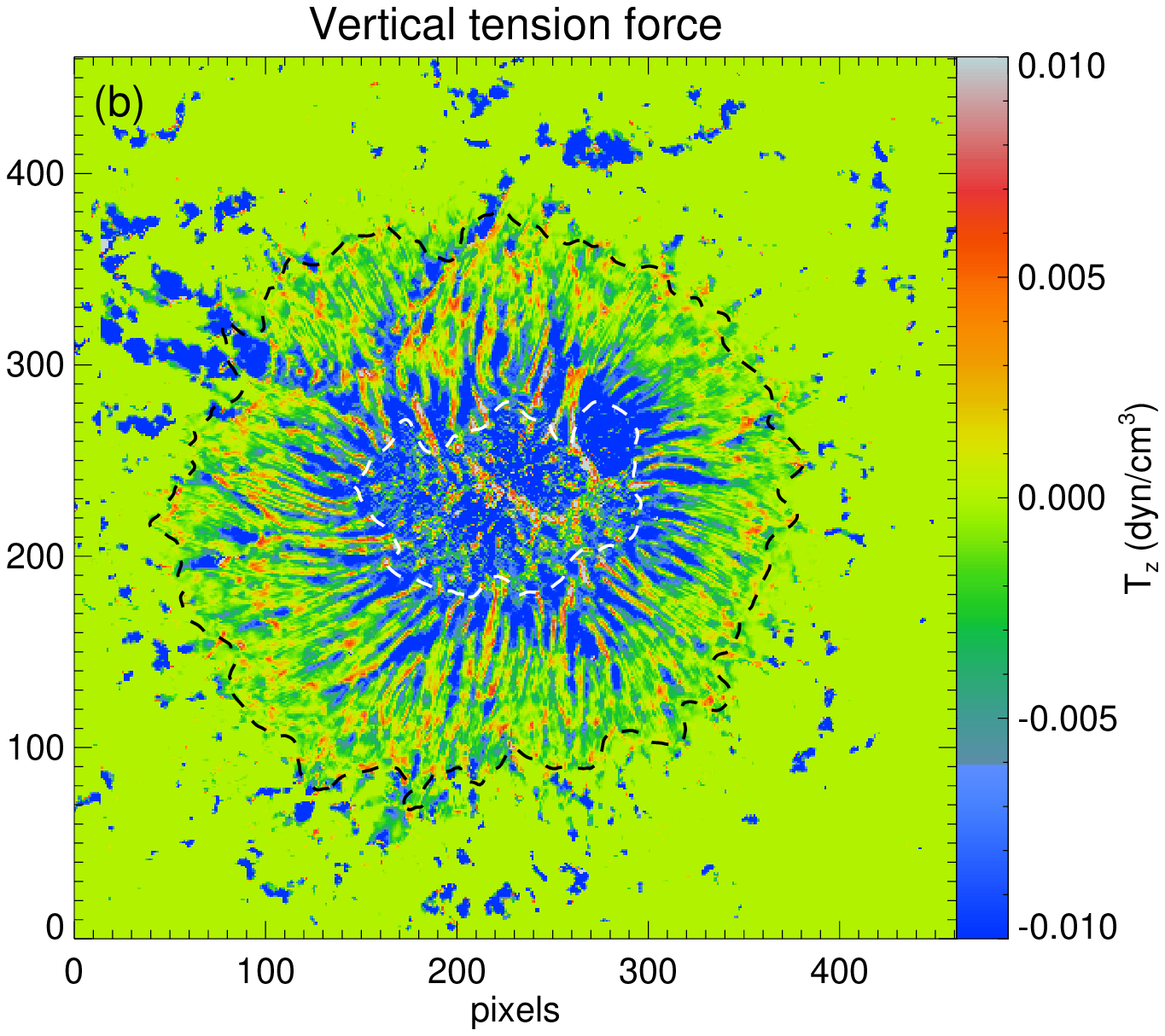}\vspace*{.5 cm}
\includegraphics[width=0.46\textwidth]{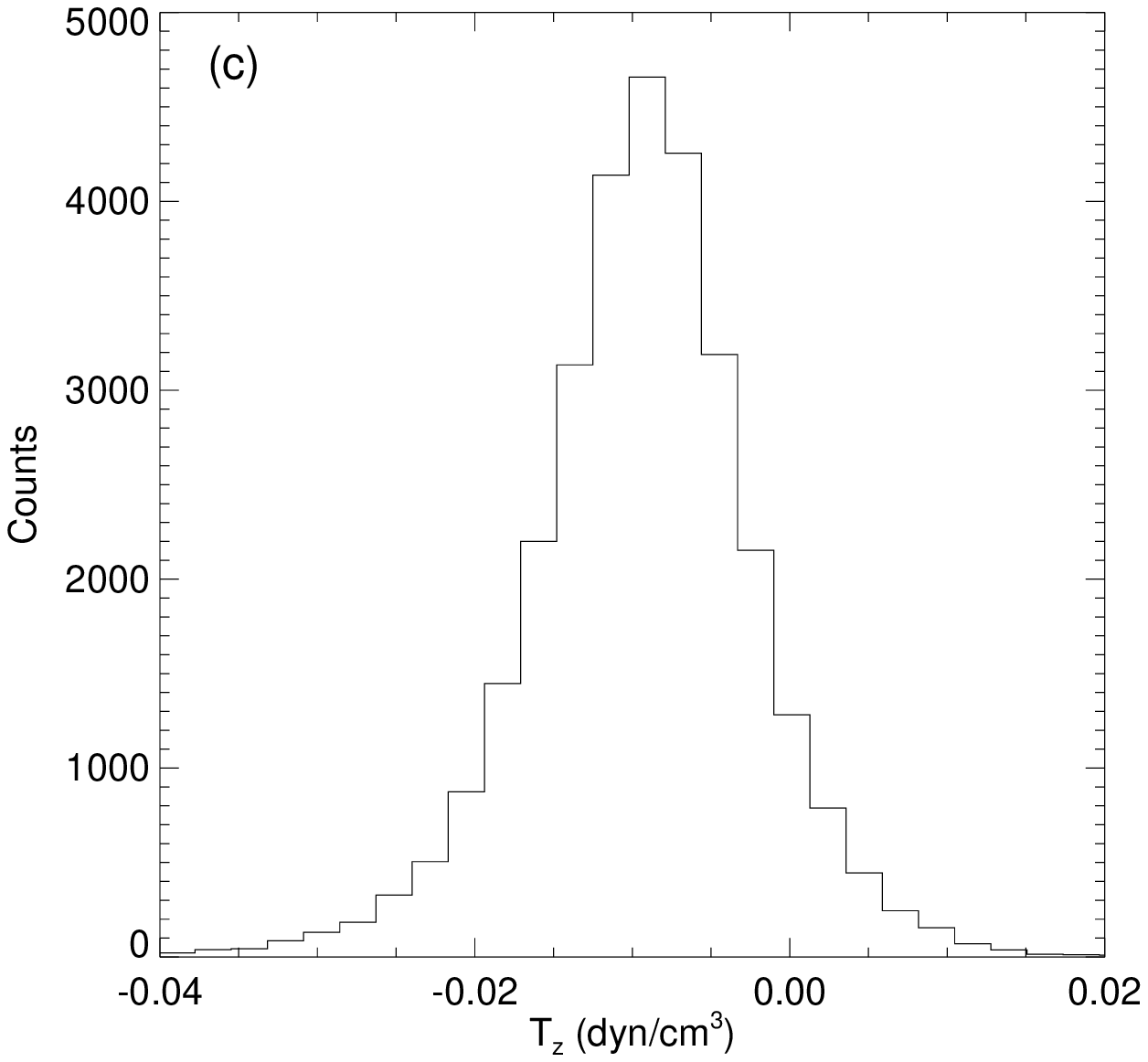}\hspace*{.2 cm}\vspace*{.5 cm}
\includegraphics[width=0.44\textwidth]{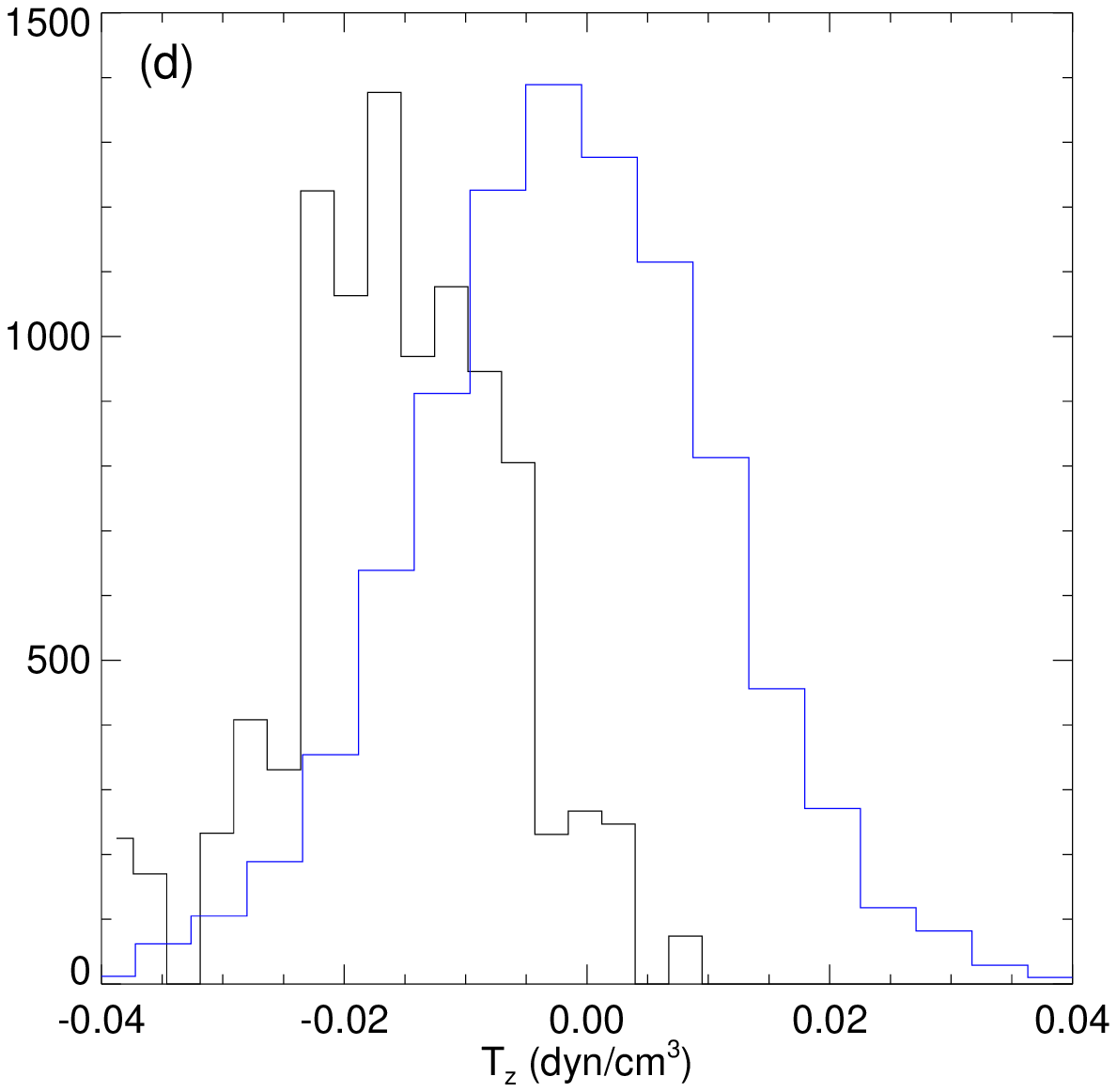}
\caption{Continuum intensity (a) and vertical tension force ($T_z$) map (b) of a typical quiet sunspot
NOAA AR 10933, observed from Hinode in normal scan mode of SOT/SP on 05 January 2007 between 12 to 13 UT.
We can note that the magnitudes of $T_z$ are high at most of the places over sunspot.
For better contrast, the magnitudes of $T_z$ ranging between -0.01 and 0.01 are
only plotted. However the histogram of full distribution of magnitudes of $T_z$
are shown in the panel (c). The white and black colored contours represent the
boundaries of the umbra and the penumbra respectively. Histograms of $T_z$ in the umbra and
penumbra are shown separately in the panel (d). The black and blue colored contours are for $T_z$ values in
the umbra and the penumbra respectively. We can infer from the peaks of the histograms that the umbral fields
are more force-free than the penumbral fields.}
\end{figure}

\begin{figure}
\epsscale{0.96}
\includegraphics[width=0.46\textwidth]{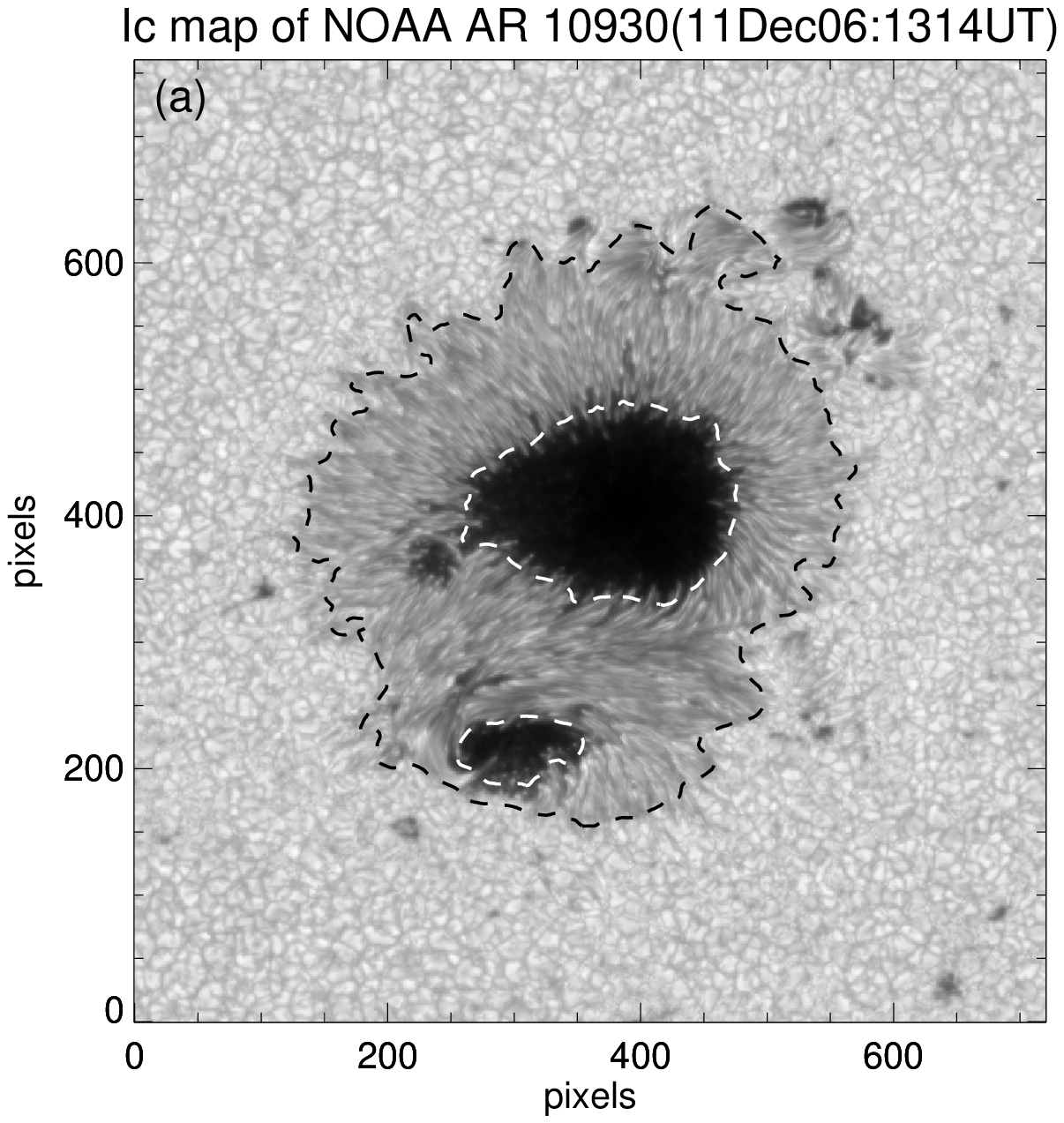}
\includegraphics[width=0.526\textwidth]{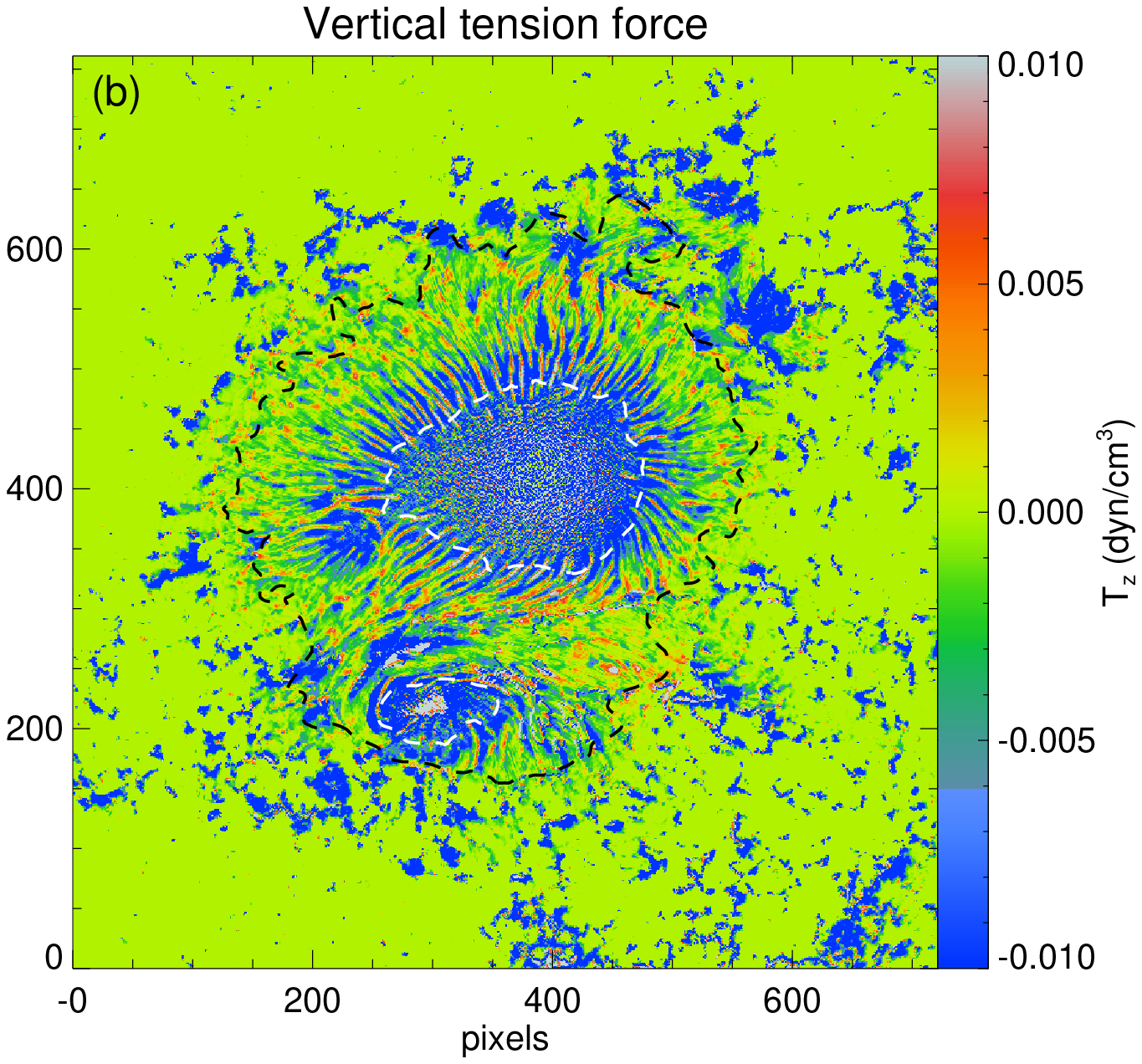}\vspace*{.5 cm}
\includegraphics[width=0.46\textwidth]{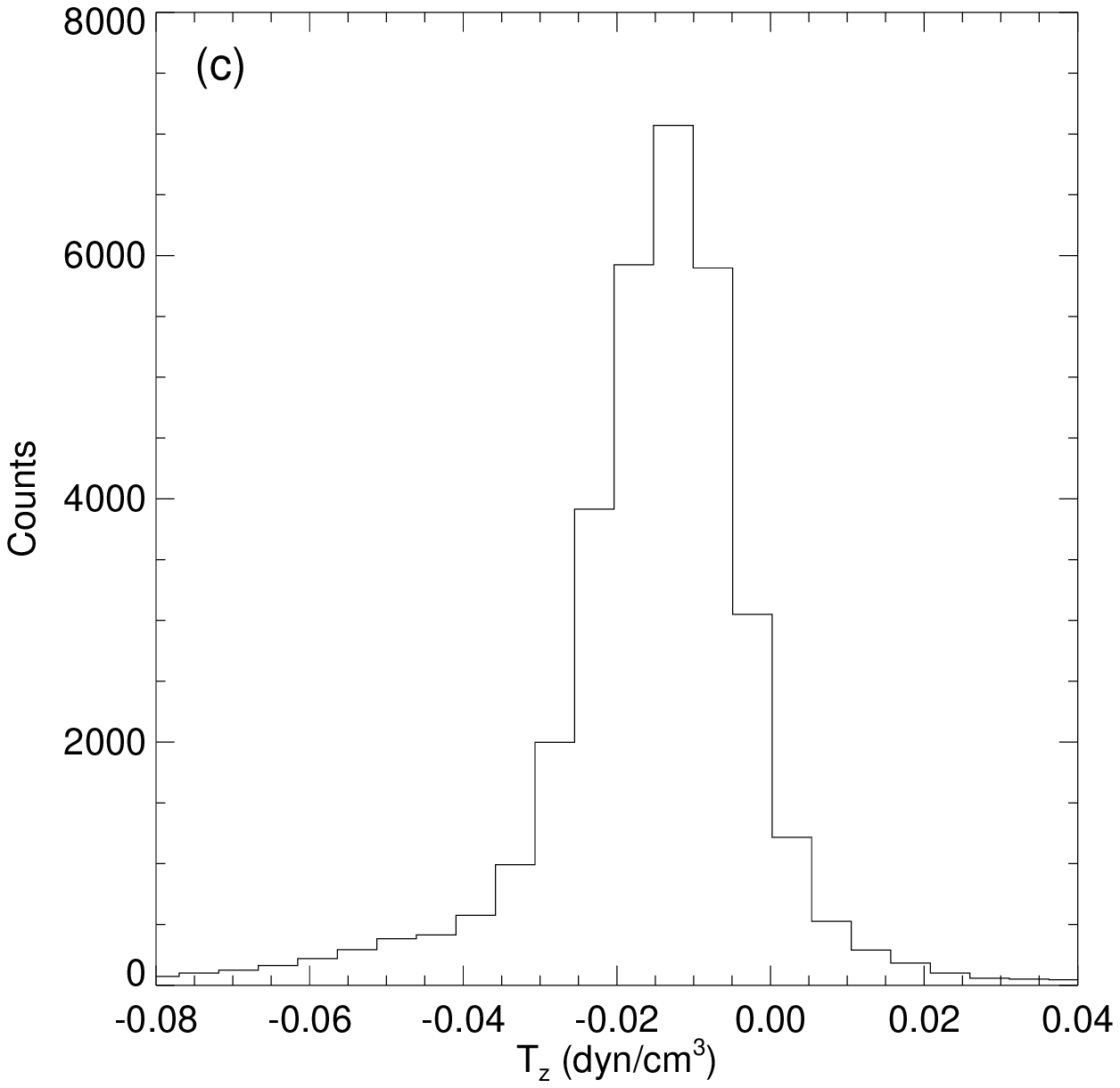}\hspace*{.2 cm}\vspace*{.5 cm}
\includegraphics[width=0.44\textwidth]{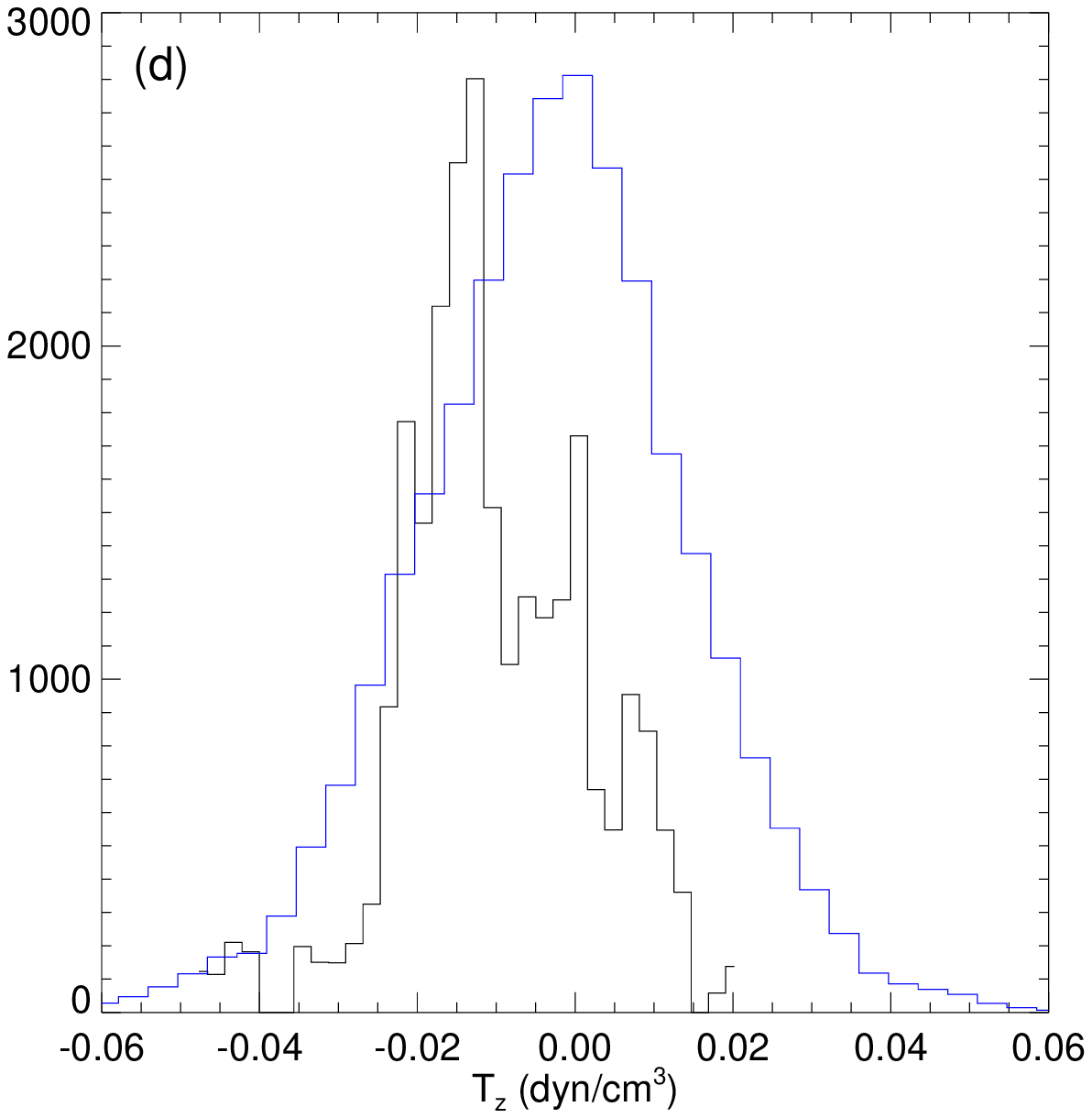}
\caption{Continuum intensity (a) and vertical tension map (b) of a typical
active/complex sunspot NOAA AR 10930, observed from Hinode (SOT/SP) on 11
December 2006 between 13 to 14 UT. We can note that the magnitudes of $T_z$
are high at most of the places over sunspot. For better contrast, the magnitudes
of $T_z$ ranging between -0.01 and 0.01 are only plotted.
However the histogram of full distribution of magnitudes of $T_z$
are shown in the panel (c). The white and black colored contours represent the
boundaries of the umbra and the penumbra respectively. Histograms of $T_z$ in the umbra and
penumbra are shown separately in the panel (d). The black and blue colored contours are for $T_z$ values in
the umbra and the penumbra respectively. We can infer from the peaks of the histograms that the
umbral fields are more force-free than the penumbral fields.}
\end{figure}

 The contours shown in Figures 1(a), 1(b), 2(a) and 2(b) distinguish between
the umbra and the penumbra of these two sunspots. The histograms of $T_z$ values
in the umbra and penumbra are plotted separately in the panels `d' of each
figure. The blue and black coloured histograms are for $T_z$ values in the penumbra
and umbra respectively. It is clearly evident from the peaks of both the
histograms that the magnitudes of $T_z$ are higher in the umbra in comparison to
the penumbra, thus indicating that the umbral fields are more force-free than the
penumbral fields. From images, we can further note that the inner penumbra also shows
high magnitudes of $T_z$, thus indicating its force-free nature.
The result that the umbral and inner penumbral fields
are more force-free in comparison to the middle and the outer penumbral fields
is in agreement with some earlier observations \citep[see, e.g.][]{sola93a,shibu04}.
The overall penumbral field seems to bear a mixed nature of magnetic fields
depending upon nature of its fine structures (penumbral fibrils).
Stronger tension forces are found in stronger and more vertical
fields \citep{venk10}. Therefore, we can also conclude that, in the penumbra,
the fields are more force-free in the spines which are more vertical and
stronger fields.

We have examined the behaviour of the tension force in all the
active regions listed in Table 1. Similar behaviour is found in all
the sunspots. The active regions at most of the places (mainly in the
umbra and in the inner penumbra) show force-free
behaviour. This serves as sufficient condition for the validity
of force-freeness of the sunspot magnetic fields. Thus the approximation
of sunspot fields as force-free magnetic fields for several purposes,
should give reliable results.

The mean and standard errors of $T_z$ are given in the Tables 1 and 2.
The standard error is computed as:
\begin{equation}
\mu_{T_z} = \frac{\sigma_{T_z}}{\sqrt{n}}
\end{equation}
where $\sigma_{T_z}$ is the standard deviation of vertical tension force
distribution and n is the total number of pixels used in the computation.
We can note that the average of T$_z$ is always negative. This is in agreement
with the magnetic flux rope models for sunspots \citep[see, e.g.][]{chit63,meye77}
which mainly dealt with the problem of global equilibrium of sunspots.

\subsection{Non-linear behaviour of the force-free fields}

Once known that the sunspot magnetic fields are nearly force-free,
we proceed to check whether the linear or non-linear
force-free field approximation will be closer to validity. For this purpose,
we examined the distribution of the force-free parameter $\alpha$ over all
the active regions that are listed in Table 1.
Two examples of $\alpha$ distribution over active regions are shown
in the left panels of Figure 3. The NOAA AR 10933 is a flare quiet and
10930 is a flare productive sunspot. Right panels show the corresponding
histograms of local $\alpha$ distribution over these active regions.
We find, unsurprisingly, that the sunspots with
very complex structures such as NOAA AR 10930 show non-linear force-free
field behaviour. Surprisingly, in the similar way, simple and quiet sunspots
like NOAA AR 10933 also show non-linear force-free field behaviour. However the
standard deviations are smaller in such simple active regions as compared
to the complex active regions. This can be noticed, as an example,
from the histograms in upper and lower right panels of Figure 3.
We can observe that the range of standard deviations in both the cases
are different, however, the distributions show similar pattern.
This leads us to believe on the validity of non-linear force-free field
approximation for both the simple and the complex sunspots.

There are several methods proposed for computation of global alpha
values \citep[see, e.g.][etc.]{pcm95,leka99,hagi04,tiw09a,tiw09e}.
Most of these methods weigh on magnetic field strength with the aim to
predict the flare activity by computing global twist of sunspots \citep{tiw09a}.
However in the present analysis, we are merely interested in the nature of linear
or non-linear distribution of $\alpha$. Thus we compute $\alpha$ locally over
an active region and take its simple mean as a global value of $\alpha$ for that AR.
The mean and standard errors of $\alpha$ for all the vector magnetograms are given in
both the Tables. The standard error in computation of $\alpha$ has been derived from:
\begin{equation}
\mu_{\alpha} = \frac{\sigma_{\alpha}}{\sqrt{n}}
\end{equation}
where $\sigma_{\alpha}$ is the standard deviation of $\alpha$
distribution and n is the total number of pixels used in the
computation. More detailed statistical investigations of distribution
of $\alpha$ can be found elsewhere \citep[see, e.g.][etc.]{pcm94,leka99,bley02,hagi04}.

\begin{figure}
\epsscale{1.}
\includegraphics[width=0.52\textwidth]{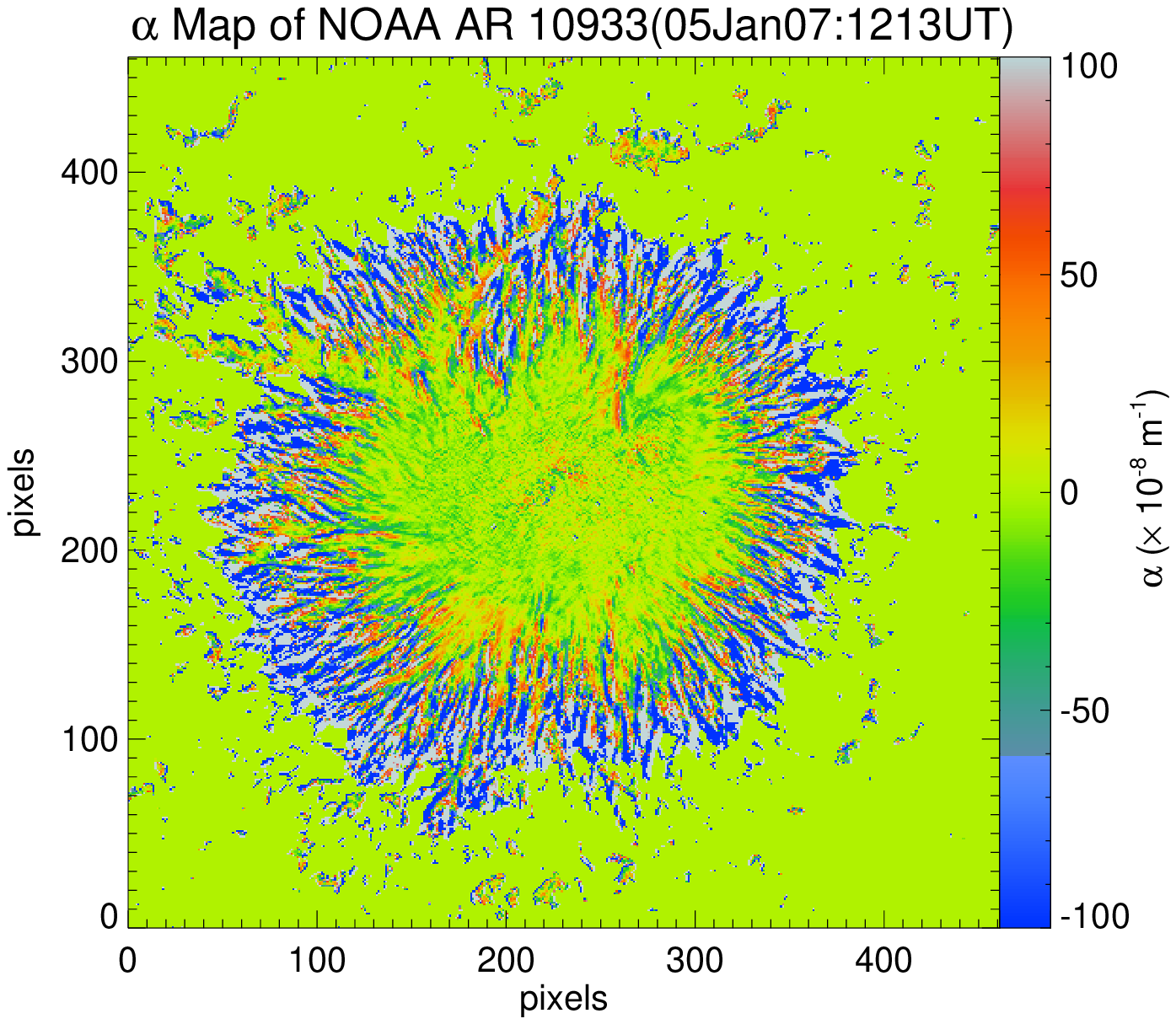} \vspace*{0.5 cm}
\includegraphics[width=0.46\textwidth]{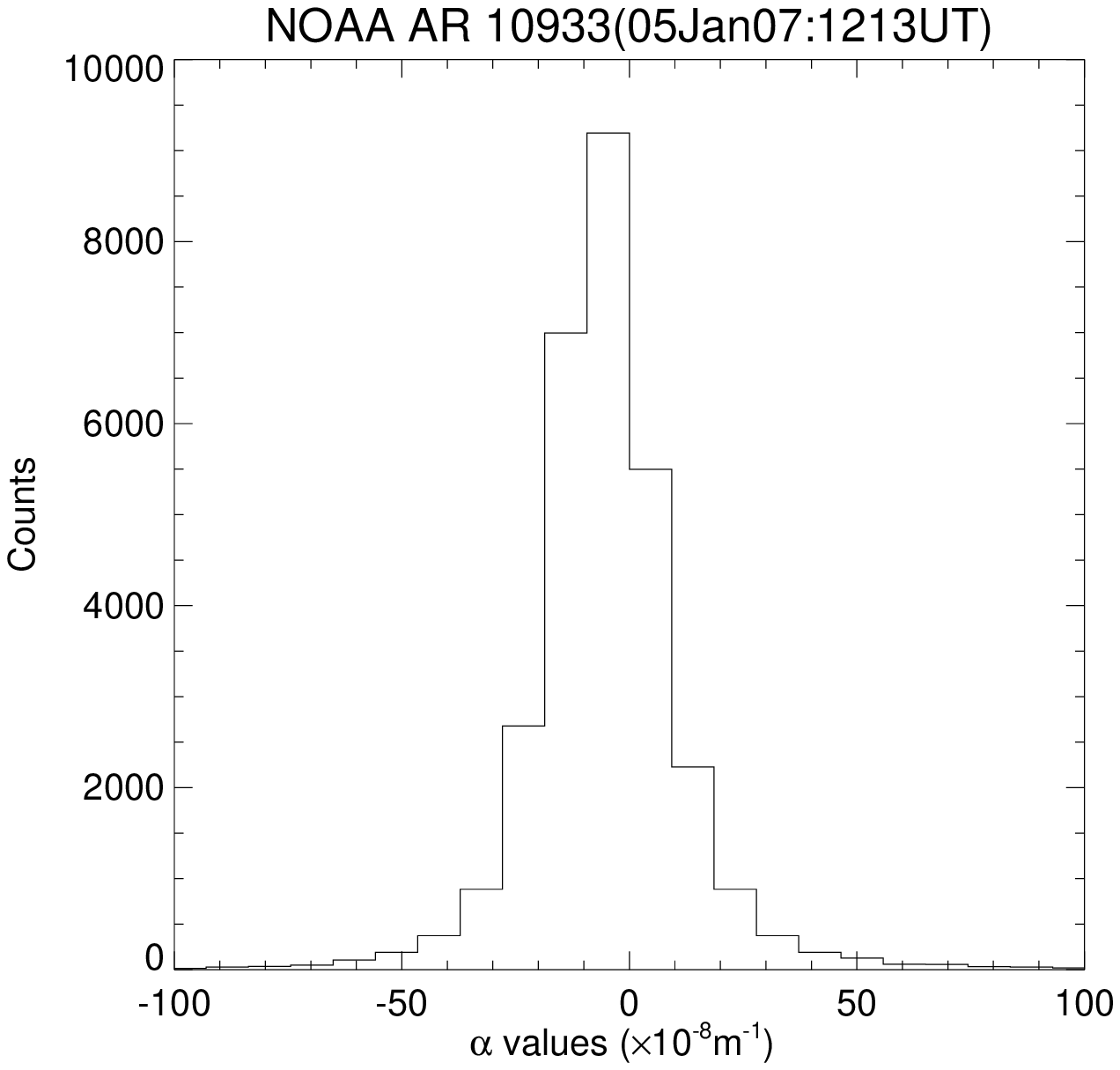}
\includegraphics[width=0.52\textwidth]{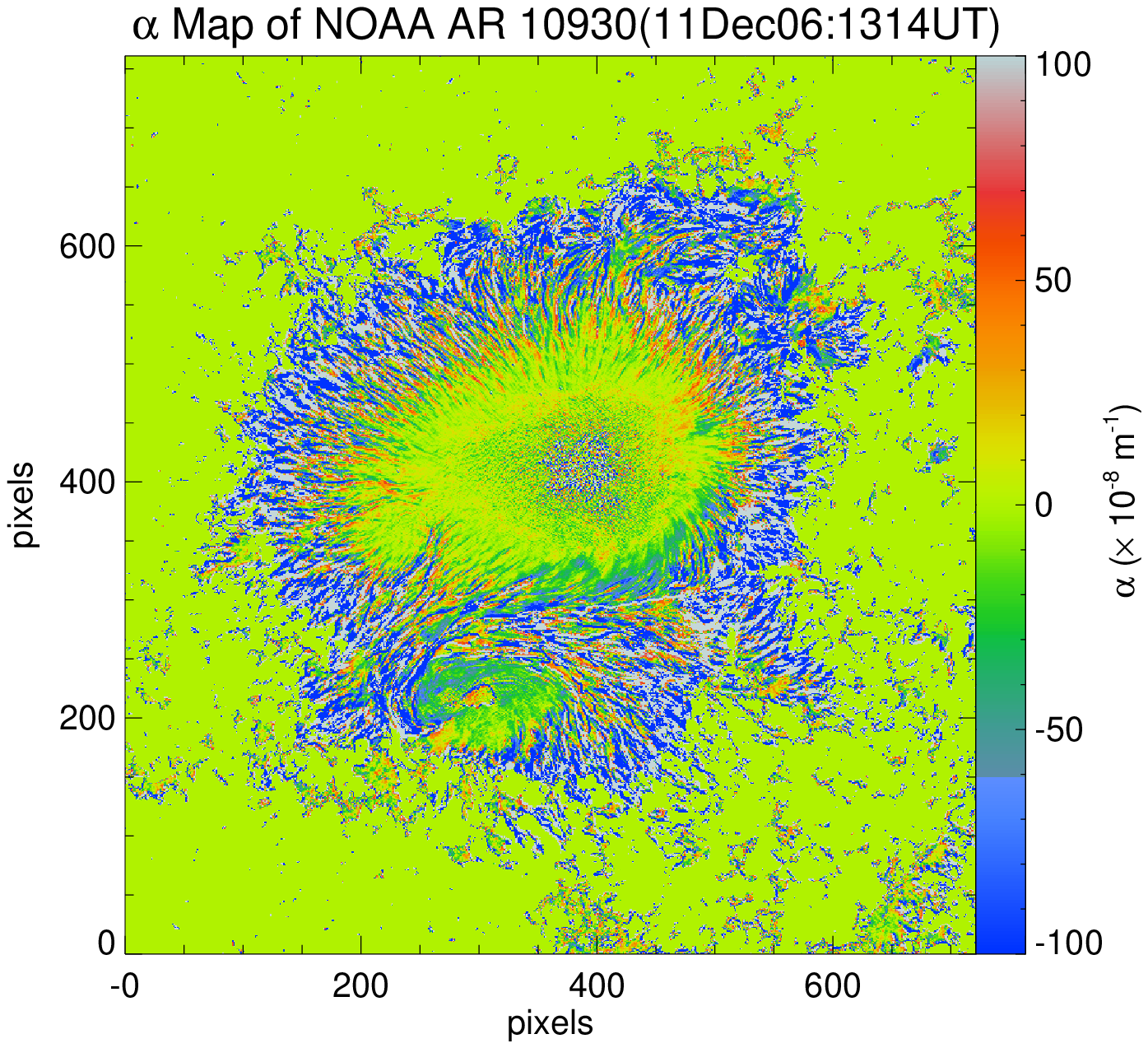} 
\includegraphics[width=0.48\textwidth]{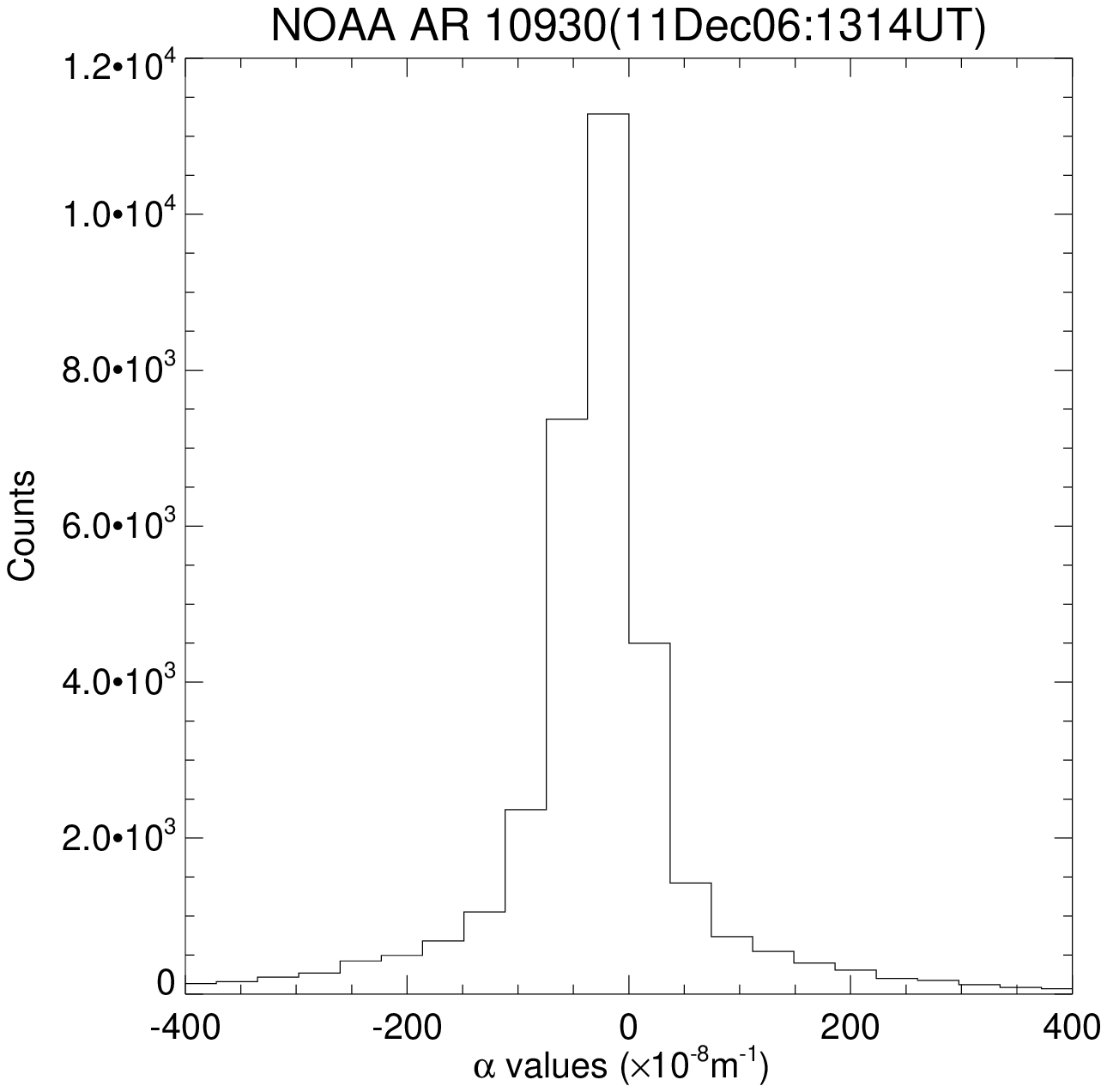}
\caption{Maps of the force-free parameter $\alpha$ for the two active regions 10933 and
10930 respectively. We can note that the $\alpha$ is distributed
non-linearly over both the sunspots.
Histograms of the $\alpha$ distribution of the NOAA ARs 10933 (quiet)
and 10930 (flare productive) are shown in right panels respectively.
We can see that the $\alpha$ is not focussed
on a very close value in either case. Although the standard deviation is approximately
three times higher in the case of highly flare productive sunspot 10930 in comparison to
the simple sunspot 10933, both the sunspots show a tendency to evince non-linear force-free
field atmosphere.}
\end{figure}

\subsection{Temporal evolution of sunspot fields}

Further, we studied temporal variation of the force-free nature of
the sunspot magnetic fields. For this purpose, we have selected 4
sunspots for which total of 60 vector magnetograms are taken
from SOT/SP. Table 2 gives details of the active regions analysed for this
purpose. The components of the net Lorentz force are plotted in Figure 4.
The results show that there are variations in the three components with time
but they are always less than unity. Most of the time, the ratios $F_x$, $F_y$
and $F_z$ are close to 0.1 and less than 0.4. However in a few cases the
magnitudes go beyond at certain stages of their temporal evolution.
In those cases also the magnitudes of $F_x$, $F_y$ and $F_z$ are
always less than 0.6. Thus, the fields never obtain a magnetic field
configuration that is much away from the force-free approximation.
The distribution of vertical tension force also shows consistency and
doesn't vary drastically with time.

From Figure 4, we note that $F_x/F_p$, $F_y/F_p$ and $F_z/F_p$ are mostly
close to zero in the cases of NOAA ARs 10930, 10933 and 10961 and never go beyond
the value of 0.4. However, it is noticeable that $F_x/F_p$, $F_y/F_p$ and $F_z/F_p$
goes up to around 0.5 at certain stages during their temporal evolution for NOAA
AR 10960. This might be indicative of some relationship between the flare
activity and the force-free field nature of sunspots. At this stage, however,
we are unable to decide on this with the limited number of the flare-productive
sunspots and in want of their time series vector magnetograms.

\begin{table}
\caption{\tiny NOAA ARs 10930, 10933, 10960 and 10961 during their temporal evolution
for few days with their observational details and results obtained.
Associated plots of $F_x/F_p$, $F_y/F_p$ and $F_z/F_p$ are shown in
Figure 4.}
\small{} \centering
\begin{tiny}
 \begin{tabular}{c c c c c c c c c}
\hline     
\hline
NOAA AR  & Date: Time(UT)   & Position  &$F_x/F_p$ &$F_y/F_p$& $F_z/F_p$ & $<\alpha>\pm\mu_{\alpha}$& $<T_z>\pm\mu_{T_z}$     & comment \\
Number   &  of Observation  &           &          &         &           & $\times10^{-8}m^{-1}$       & $\times10^{-3}dyn~cm^{-3}$ &     \\
\hline
10930    & 10 Dec 2006: 0102  & S04E14(t)&   0.078  &  -0.003  & -0.069  & $-2.854 \pm 0.719$     & $-6.501 \pm 0.061$   & complex, rotating   \\
10930    & 10 Dec 2006: 2122  & S04E03   &   0.115  &  0.022   & -0.051  & $-5.691 \pm 0.599$     & $-7.509 \pm 0.048$   & complex, rotating    \\
10930    & 11 Dec 2006: 0304  & S04W01   &   0.079  &  0.029   & -0.076  & $-7.875 \pm 0.536$     & $-7.267 \pm 0.045$   & complex, rotating    \\
10930    & 11 Dec 2006: 0809  & S04W04   &   0.059  &  0.033   & -0.089  & $-6.946 \pm 0.508$     & $-6.213 \pm 0.044$   & complex, rotating    \\
10930    & 11 Dec 2006: 1112  & S04W06   &   0.047  &  0.034   & -0.086  & $-10.252 \pm 0.534$    & $-6.462 \pm 0.048$   & complex, rotating    \\
10930    & 11 Dec 2006: 1314  & S05W07   &   0.039  &  0.038   & -0.062  & $-8.604 \pm 0.316$     & $-5.749 \pm 0.028$   & complex, rotating    \\
10930    & 11 Dec 2006: 1718  & S05W08   &   0.016  &  0.044   & -0.077  & $-9.617 \pm 0.562$     & $-6.535 \pm 0.050$   & complex, rotating    \\
10930    & 11 Dec 2006: 2021  & S05W10   &  -0.001  &  0.047   & -0.088  & $-10.035 \pm 0.555$    & $-6.617 \pm 0.051$   & complex, rotating    \\
10930    & 11 Dec 2006: 2324  & S05W12   &  -0.013  &  0.045   & -0.063  & $-10.490 \pm 0.595$    & $-6.401 \pm 0.059$   & complex, rotating    \\
10930    & 12 Dec 2006: 0304  & S06W14(t)&   0.099  &  0.039   & -0.029  & $-13.988 \pm 0.599$    & $-5.959 \pm 0.069$   & complex, rotating    \\
10930    & 12 Dec 2006: 1011  & S06W18(t)&   0.102  &  0.052   & -0.024  & $-11.561 \pm 0.325$    & $-4.971 \pm 0.047$   & complex, rotating    \\
10930    & 12 Dec 2006: 1516  & S06W21(t)&   0.119  & -0.060   & -0.042  & $-8.919 \pm 0.547$     & $-5.261 \pm 0.056$   & complex, rotating    \\
10930    & 12 Dec 2006: 1718  & S06W22(t)&   0.118  & -0.063   & -0.044  & $-8.342 \pm 0.509$     & $-5.359 \pm 0.053$   & complex, rotating    \\
10930    & 12 Dec 2006: 2021  & S06W24(t)&   0.127  &  0.065   & -0.047  & $-8.803 \pm 0.539$    & $-5.697 \pm 0.063$   & complex, rotating   \\
10930    & 13 Dec 2006: 0405  & S06W29(t)&   0.137  &  0.078   & -0.025  & $-11.117 \pm 0.551$    & $-6.389 \pm 0.069$   & complex, rotating   \\
10930    & 13 Dec 2006: 0708  & S06W31(t)&   0.136  &  0.079   & -0.002  & $-9.793 \pm 0.617$    & $-6.153 \pm 0.076$   & complex, rotating   \\
10930    & 13 Dec 2006: 1314  & S06W34(t)&   0.125  &  0.083   & -0.007  & $-6.289 \pm 0.302$     & $-5.001 \pm 0.034$   & complex, rotating   \\
10930    & 13 Dec 2006: 1617  & S06W35(t)&   0.126  &  0.084   & -0.006  & $-7.712 \pm 0.555$     & $-6.061 \pm 0.060$   & complex, rotating   \\
10933    & 04 Jan 2007: 1112  & S04E19(t)&   0.202  &   0.029  & -0.102  & $-1.082 \pm 0.181$     & $-6.255 \pm 0.026$   & quiet   \\
10933    & 04 Jan 2007: 1819  & S04E15(t)&  -0.049  &   0.018  & -0.174  & $-3.019 \pm 0.308$     & $-6.110 \pm 0.047$   & quiet   \\
10933    & 05 Jan 2007: 0001  & S04E11   &   0.125  &   0.017  & -0.146  & $-1.254 \pm 0.233$     & $-6.721 \pm 0.050$   & quiet   \\
10933    & 05 Jan 2007: 0708  & S04E06   &   0.085  &   0.003  & -0.113  & $-0.569 \pm 0.233$     & $-6.624 \pm 0.043$   & quiet   \\
10933    & 05 Jan 2007: 1213  & S04E03   &   0.045  &   0.002  & -0.108  & $-0.557 \pm 0.123$     & $-6.341 \pm 0.026$   & quiet   \\
10933    & 06 Jan 2007: 0001  & S03W03   &  -0.032  &  -0.017  & -0.134  & $-0.520 \pm 0.170$     & $-5.765 \pm 0.038$   & quiet   \\
10933    & 06 Jan 2007: 0405  & S03W06   &  -0.065  &  -0.018  & -0.128  & $-0.406 \pm 0.120$     & $-5.650 \pm 0.025$   & quiet   \\
10933    & 06 Jan 2007: 1000  & S03W09   &  -0.106  &   0.001  & -0.112  & $-0.847 \pm 0.112$     & $-6.247 \pm 0.028$   & quiet   \\
10933    & 06 Jan 2007: 1213  & S03W10   &  -0.133  &   0.002  & -0.095  & $-0.458 \pm 0.132$     & $-6.457 \pm 0.030$   & quiet   \\
10933    & 07 Jan 2007: 0001  & S04W17(t)&   0.135  &  -0.005  & -0.158  & $-2.223 \pm 0.217$     & $-5.175 \pm 0.036$   & quiet   \\
10933    & 07 Jan 2007: 1213  & S04W20(t)&   0.103  &  -0.003  & -0.131  & $-2.064 \pm 0.261$     & $-4.524 \pm 0.036$   & quiet   \\
10933    & 07 Jan 2007: 1819  & S04W25(t)&   0.151  &  -0.018  & -0.072  & $-2.382 \pm 0.257$     & $-5.108 \pm 0.037$   & quiet    \\
10960    & 05 Jun 2007: 0607  & S08E29(t)&   0.079  &  0.065   & -0.144  & $-3.270 \pm 0.610$     & $-4.188 \pm 0.049$   & complex, active   \\
10960    & 05 Jun 2007: 1112  & S08E26(t)&   0.087  &  0.079   & -0.222  & $-2.755 \pm 0.439$     & $-3.773 \pm 0.037$   & complex, active   \\
10960    & 06 Jun 2007: 0203  & S07E19(t)&   0.119  &  0.074   & -0.335  & $-3.423 \pm 0.480$     & $-4.465 \pm 0.046$   & complex, active   \\
10960    & 06 Jun 2007: 0708  & S07E17(t)&   0.126  &  0.076   & -0.375  & $-2.711 \pm 0.490$     & $-4.525 \pm 0.048$   & complex, active   \\
10960    & 06 Jun 2007: 1213  & S07E14(t)&   0.111  &  0.089   & -0.441  & $-2.226 \pm 0.374$     & $-4.394 \pm 0.038$   & complex, active   \\
10960    & 06 Jun 2007: 1920  & S07E10   &   0.192  &  0.086   & -0.464  & $-1.252 \pm 0.373$     & $-4.857 \pm 0.037$   & complex, active   \\
10960    & 06 Jun 2007: 2223  & S07E09   &   0.198  &  0.092   & -0.451  & $-1.105 \pm 0.478$     & $-5.678 \pm 0.048$   & complex, active   \\
10960    & 07 Jun 2007: 0304  & S07E07   &   0.137  &  0.093   & -0.482  & $-1.588 \pm 0.446$     & $-5.625 \pm 0.045$   & complex, active   \\
10960    & 07 Jun 2007: 1920  & S07W03   &   0.008  &  0.109   & -0.535  & $-2.302 \pm 0.501$     & $-6.127 \pm 0.050$   & complex, active   \\
10960    & 08 Jun 2007: 0708  & S07W09   &  -0.178  &  0.100   & -0.514  & $-1.631 \pm 0.566$     & $-6.821 \pm 0.055$   & complex, active   \\
10960    & 08 Jun 2007: 0809  & S07W10   &  -0.204  &  0.103   & -0.505  & $-1.055 \pm 0.570$     & $-6.928 \pm 0.055$   & complex, active   \\
10960    & 08 Jun 2007: 1314  & S07W12(t)&  -0.223  &  0.129   & -0.553  & $0.302 \pm 0.418$      & $-6.101 \pm 0.039$   & complex, active   \\
10960    & 08 Jun 2007: 1516  & S07W14(t)&   0.225  &  0.085   & -0.559  & $-2.764 \pm 0.360$     & $-5.898 \pm 0.034$   & complex, active   \\
10960    & 08 Jun 2007: 1819  & S07W16(t)&  -0.238  &  0.079   & -0.533  & $-4.118 \pm 0.395$     & $-6.120 \pm 0.036$   & complex, active   \\
10960    & 08 Jun 2007: 2021  & S07W17(t)&  -0.208  &  0.073   & -0.519  & $-4.120 \pm 0.423$     & $-6.233 \pm 0.042$   & complex, active   \\
10961    & 29 Jun 2007: 0100  & S11E32(t)&  -0.053  &  0.147   & -0.092  & $-4.429 \pm 0.794$     & $-3.592 \pm 0.054$   & quiet    \\
10961    & 29 Jun 2007: 0700  & S11E28(t)&  -0.057  &  0.147   & -0.126  & $-5.473 \pm 0.819$     & $-4.023 \pm 0.055$   & quiet    \\
10961    & 29 Jun 2007: 1819  & S11E21(t)&  -0.026  &  0.144   & -0.176  & $-4.908 \pm 0.875$     & $-4.384 \pm 0.061$   & quiet    \\
10961    & 30 Jun 2007: 0000  & S12E19(t)&  -0.032  &  0.137   & -0.182  & $-4.654 \pm 0.972$     & $-4.569 \pm 0.073$   & quiet    \\
10961    & 30 Jun 2007: 0405  & S12E17(t)&  -0.044  &  0.149   & -0.217  & $-6.521 \pm 0.863$     & $-4.661 \pm 0.062$   & quiet    \\
10961    & 30 Jun 2007: 0910  & S12E14(t)&  -0.025  &  0.157   & -0.249  & $-5.359 \pm 0.973$     & $-5.575 \pm 0.078$   & quiet    \\
10961    & 30 Jun 2007: 1100  & S13E13(t)&  -0.002  &  0.167   & -0.307  & $-4.725 \pm 0.515$     & $-4.561 \pm 0.039$   & quiet    \\
10961    & 30 Jun 2007: 2223  & S13E08(t)&  -0.010  &  0.128   & -0.174  & $-4.235 \pm 1.549$     & $-5.579 \pm 0.116$   & quiet    \\
10961    & 01 Jul 2007: 0304  & S13E05(t)&   0.024  &  0.172   & -0.271  & $-4.754 \pm 0.939$     & $-5.753 \pm 0.072$   & quiet    \\
10961    & 01 Jul 2007: 0809  & S13E02(t)&  -0.047  &  0.183   & -0.257  & $-5.437 \pm 1.064$     & $-6.404 \pm 0.082$   & quiet    \\
10961    & 01 Jul 2007: 1314  & S13W01(t)&  -0.054  &  0.186   & -0.269  & $-2.786 \pm 0.954$     & $-5.879 \pm 0.073$   & quiet    \\
10961    & 01 Jul 2007: 2000  & S12W05(t)&   0.075  &  0.180   & -0.279  & $-2.259 \pm 0.889$     & $-5.545 \pm 0.069$   & quiet    \\
10961    & 02 Jul 2007: 0708  & S10W12(t)&   0.125  &  0.169   & -0.298  & $-2.423 \pm 0.934$     & $-6.510 \pm 0.072$   & quiet    \\
10961    & 02 Jul 2007: 1200  & S10W14(t)&   0.169  &  0.163   & -0.278  & $-2.387 \pm 0.953$     & $-6.209 \pm 0.074$   & quiet    \\
10961    & 02 Jul 2007: 1900  & S10W18(t)&   0.160  &  0.088   & -0.190  & $-4.776 \pm 1.363$     & $-4.905 \pm 0.098$   & quiet    \\
\hline
(t) :  \it transformed \\
\end{tabular}
\end{tiny}
\end{table}

\begin{figure}
\epsscale{0.82}
\plotone{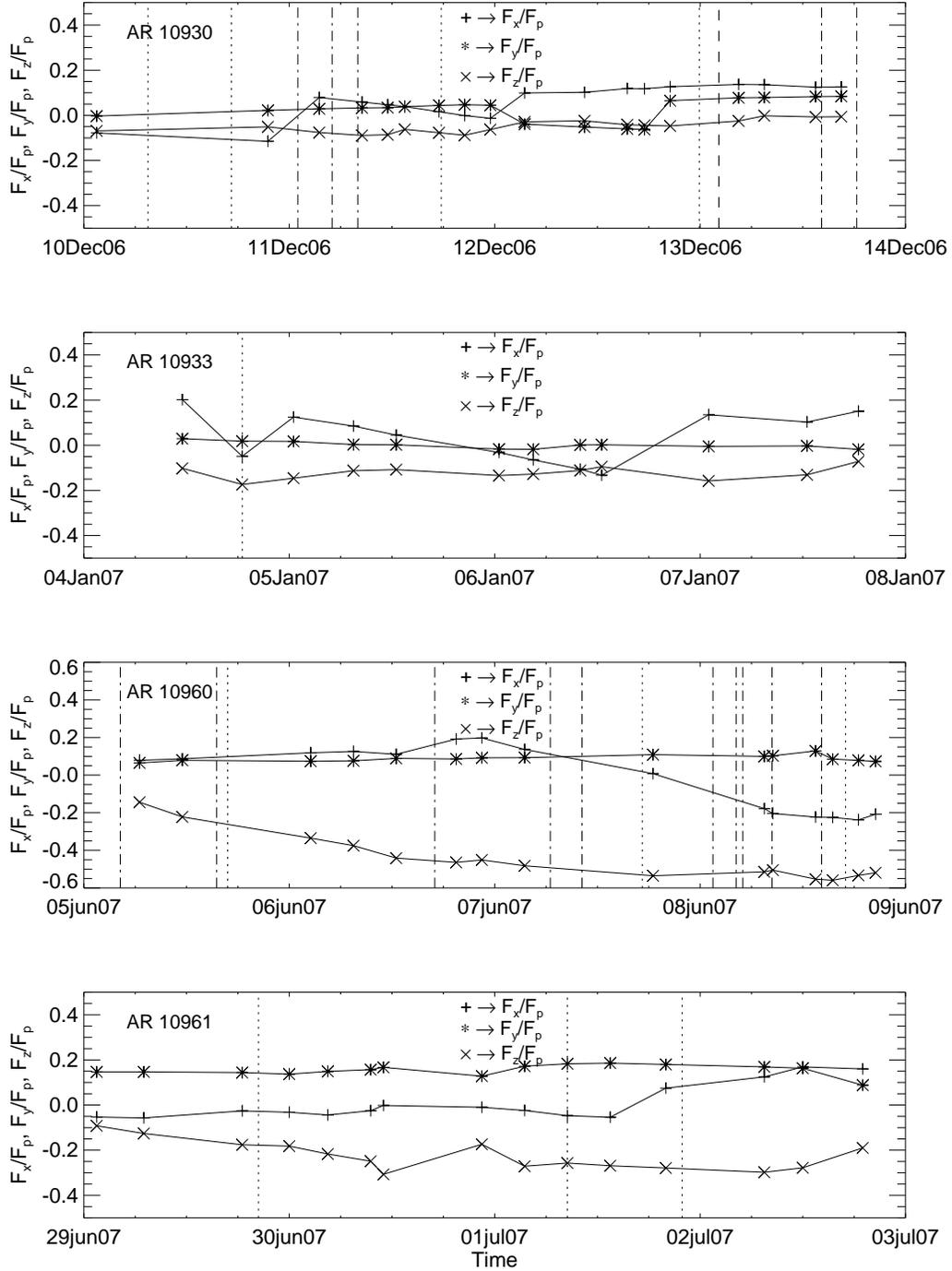}
\caption{Plots of temporal evolution of $F_x/F_p$, $F_y/F_p$ and $F_z/F_p$ for
NOAA ARs 10930, 10933, 10960 and 10961 as given in Table 2. We can note that the magnitudes are
always less than unity, mostly close to 0.1 and never go beyond 0.4 except for AR 10960, where it
goes up to 0.55. The vertical dotted lines
represent timings of B-class flares, dashed-dotted lines represent timings of C-class flares and
dashed line (only on Dec 13 2006) represents X-class flare. We can see that, although
$F_x/F_p$, $F_y/F_p$ and $F_z/F_p$ depend on the time evolution, they don't show a systematic
relationship with the flare activity.
}
\end{figure}

The vertical lines in the plots of Figure 4 represent the timings
of the associated flares. The dashed line represents the timing of X-class
flare (only one on 13 Dec 2006), dashed-dotted lines represent timings of
the C-class flares and dotted lines represent timings of the B-class flares.
As can be noted from the plots that, even if $F_x/F_p$, $F_y/F_p$ and $F_z/F_p$
depend on the evolutionary stages of the sunspot magnetic fields,
we do not find any systematic relationship between the force-freeness of
the sunspot fields and the associated flare activity.

The computation of field gradients might be influenced by the optical
depth corrugation effects. An estimation however shows that the relative
variation in the field strength and thus in the tension force will be 10\%,
which is much smaller than the observed spatial variation of vertical tension
force \citep{venk10}.
Acute angle method for azimuthal ambiguity resolution works well whenever
transverse magnetic field vector makes an angle of less than 90 degrees
with potential transverse field. Most of the active regions studied are
simple sunspots with no complex polarity inversion lines except in the
NOAA AR 10930. Thus, ambiguity is resolved properly in most of the active
regions. We are interested in the magnitude of vertical tension force
in the present study and not in the sign of it, thus we can ignore the
influence of ambiguity disambiguation at neutral lines whenever present.
Another possibility of error is because of the large
scanning time of Hinode (SOT/SP). If the evolution of magnetic field
is faster than the scan time of SP then this might cause an error
in the results. This is unlikely because of larger evolution
time of sunspot magnetic fields, unless the sunspot is involved in an
ongoing flare. For an example if we consider 60 arcsec size of an active
region and Alfv$\acute{e}$n speed of about 10 km/sec on the photosphere,
then the Alfv$\acute{e}$n travel time across the active region will be
of the order of 1 hour which is not smaller than the scan time of SP for
such field of view.

\section{Discussion and Conclusions}

Using high spatial resolution spectro-polarimetric data obtained
from SOT/SP aboard the Hinode, we find that the photospheric
sunspot magnetic fields can be approximated to the force-free
field configurations.
The necessary and the sufficient, both conditions dealing with global
and local nature of sunspot magnetic fields respectively, indicate that the
sunspot magnetic fields are nearly force-free. This result is consistent with
the results obtained by \cite{moon02}. However, \cite{moon02} verified
the force-free approximation by using only necessary conditions
given by inequalities in Equation 10. We have also investigated the spatial
distribution of vertical tension force which serves as sufficient condition.
After examining the distribution of the magnitude of vertical tension force
over all the sunspots listed in Table 1, we conclude that the magnetic fields
in sunspot fine structures over most of the sunspot area are nearly
force-free. We, further, distinguish between the nature of the umbral and
the penumbral magnetic fields. We find that the umbral fields are more
force-free than the penumbral fields. We also find that over penumbra, inner
penumbra are more force free than the middle and outer penumbra. From the
distribution of vertical tension force ($T_z$), it is found that the stronger
and more vertical fields (spines) exhibit higher magnitudes of $T_z$ (cf.
Figure 2 of \cite{venk10}). Thus, we can also conclude that the spines are more
force-free in the penumbra.

Firstly, a study with similar purpose was performed
by \cite{metc95} using observations of the Stokes parameters at six
wavelengths within the spectral lines of Na I $\lambda$5858 \AA\
using derivative method. They inspected the nature of Lorentz
forces by using the necessary conditions.
Their results that the magnetic field is
not force-free in the photosphere and becomes force-free above
certain heights in the lower chromosphere has been subjected to an
underestimation of transverse field due to their weak-field approximation
\citep{moon02}.

The calculations of vertical current density and alpha parameter over
sunspots observed from SOT/SP has been performed recently by
\cite{su09,tiw09b,tiw09e}. The distribution of $\alpha$ is found
to be non-uniform over sunspots.
After verifying that the sunspot fields are almost force-free,
we have examined the range of the distribution of alpha over
several sunspots, including simple and complex active regions,
to inspect the imminence of linear or non-linear force-free field approximation.
Our results show that the non-linear force-free field approximation is
applicable not only in complex active regions such as NOAA AR 10930 but also
in the simple active regions such as NOAA AR 10933 (see Figure 3).
This result is not in accordance with that of
\cite{moon02}. The reason could be the difference in the quality of data used.
The resolution of data used by \cite{moon02} might have not been sufficient
to obtain the non-linear behaviour of alpha over sunspots they studied.
Thus, the investigations using high resolution data lead us to
believe that the sunspot magnetic fields are nearly force-free
in nature and they imitate non-linear force-free behaviour.

A study of temporal evolution of Lorentz force components in
60 vector magnetograms of four active regions
(see Table 2 and Figure 4) shows variations in the Lorentz force
components normalized by F$_p$ i.e., $|F_x/F_p|$,
$|F_y/F_p|$ and $|F_z/F_p|$. They are always
lesser than unity, mostly closer to 0.1 and never show higher values
greater than 0.6, thus remaining under the validity of the force-free
field approximation. These sunspots, in every phase of their evolution,
show immediacy with the force-free approximation of their magnetic
fields. From Figure 4, we further tried to find any relationship
between the temporal evolution of the force-free field nature of
sunspots and the associated flare-activity. Although $|F_x/F_p|$,
$|F_y/F_p|$ and $|F_z/F_p|$ depend on the evolutionary stages of
active regions, they don't show a systematic relationship with
the occurrence of flares associated with those active regions.

We depend on the extrapolations of photospheric magnetic
fields to obtain the coronal vector magnetic fields owing to
inadequacy of coronal field measurements. As it is well known and
also discussed in the introduction, that all the extrapolation
techniques have to rely on the photospheric vector field measurements
and also on the force-free approximation of these magnetic fields.
The effect of several possible inaccuracies in the vector field measurement
on the non-linear force-free field modelings has recently been studied by
\cite{derosa09,wieg10}. Coronal magnetic field reconstruction by
extrapolations of photospheric magnetic fields under non-linear
modelings have shown satisfactory results by nearly matching with
the coronal observations \citep{mccl94,bley02,wieg05,schr06,metc08}.
These results then also support our conclusion that the sunspot
magnetic fields are close to satisfy the non-linear force-free field
approximation.

A preprocessing procedure has been developed by \cite{wieg06} to
make the observed data completely force-free as a suitable boundary
conditions for force-free extrapolations. This has led more reliable
better extrapolations of the fields. The method used by
\cite{wieg06} however uses only necessary condition to make the fields
force-free. Also, most of the time observed field of view suffers
from the flux imbalance leading to an inaccuracy in the preprocessing.
The preprocessing might be improved by using the
sufficient condition (as described in Section 2.2) in preparing the
boundary data for the force-free field extrapolations.

There is, however, good enough scope of improvement of the present study.
At present our investigation is limited to computation of vertical component
of the magnetic tension force. Whereas in a recent study by \cite{pusc10},
it has been found that the horizontal component of Lorentz force is greater
than the vertical component. Detailed analysis of full sunspot magnetic fields
within certain optical heights are required to understand full scenario of
force-free nature of the sunspot magnetic fields. In a forthcoming work,
we plan to look into all the three components of the equilibrium forces in
the height dependent inversions of a complete sunspot.

\acknowledgments {\bf Acknowledgements}\\

I thank the referee for very constructive comments which improved the
manuscript significantly. I am indebted to Professors P. Venkatakrishnan
and B. C. Low for their valuable suggestions and comments for improvement of
the manuscript, and Professor E. N. Parker for reading an early version of
the manuscript. I thank Professor S. K. Solanki for useful discussions on
magnetic and thermal energies. I would also like to thank to all the members
of Solar Lower Atmosphere and Magnetism (SLAM) group at MPS for their useful
comments during a SLAM group meeting.
Some of the Hinode SOT/SP Inversions were conducted at NCAR under the
framework of the Community Spectro-polarimetric Analysis Center
(CSAC; \url{http://www.csac.hao.ucar.edu/}). The help provided by Professor Bruce
Lites in handling CSAC data is sincerely acknowledged.
Hinode is a Japanese mission developed and launched by ISAS/JAXA,
collaborating with NAOJ as a domestic partner, NASA and STFC (UK) as
international partners. Scientific operation of the Hinode mission is
conducted by the Hinode science team organized at ISAS/JAXA.
This team mainly consists of scientists from institutes in the partner
countries. Support for the post-launch operation is provided by JAXA and NAOJ
(Japan), STFC (U.K.), NASA (U.S.A.), ESA, and NSC (Norway).


\begin{thebibliography}{}
%
\bibitem[\protect\citeauthoryear{{Aly}}{{Aly}}{1984}]{aly84}
{Aly}, J.~J. 1984, \apj, 283, 349
%
\bibitem[\protect\citeauthoryear{{Amari} et~al.}{{Amari} et~al.}{1997}]{amar97}
{Amari}, T., {Aly}, J.~J., {Luciani}, J.~F., {Boulmezaoud}, T.~Z.,  \& {Mikic},
  Z. 1997, \solphys, 174, 129
%
\bibitem[\protect\citeauthoryear{{Amari}, {Boulmezaoud}, \& {Aly}}{{Amari}
  et~al.}{2006}]{amar06}
{Amari}, T., {Boulmezaoud}, T.~Z.,  \& {Aly}, J.~J. 2006, \aap, 446, 691
%
\bibitem[\protect\citeauthoryear{{Amari}, {Boulmezaoud}, \& {Mikic}}{{Amari}
  et~al.}{1999}]{amar99}
{Amari}, T., {Boulmezaoud}, T.~Z.,  \& {Mikic}, Z. 1999, \aap, 350, 1051
%
\bibitem[\protect\citeauthoryear{{Bleybel} et~al.}{{Bleybel}
 et~al.}{2002}]{bley02}
{Bleybel}, A., {Amari}, T., {van Driel-Gesztelyi}, L.,  \& {Leka}, K.~D. 2002,
  \aap, 395, 685
%
\bibitem[\protect\citeauthoryear{{Borrero}, {Lites}, \& {Solanki}}{{Borrero}
  et~al.}{2008}]{borr08}
{Borrero}, J.~M., {Lites}, B.~W.,  \& {Solanki}, S.~K. 2008, \aap, 481, L13
%
\bibitem[\protect\citeauthoryear{{Chandrasekhar}}{{Chandrasekhar}}{1956a}]{chandra56b}
{Chandrasekhar}, S. 1956a, \apj, 124, 232
%
\bibitem[\protect\citeauthoryear{{Chandrasekhar}}{{Chandrasekhar}}{1956b}]{chandra56a}
{Chandrasekhar}, S. 1956b, Proceedings of the National Academy of Science, 42,
  1
%
\bibitem[\protect\citeauthoryear{{Chandrasekhar}}{{Chandrasekhar}}{1961}]{chandra61}
{Chandrasekhar}, S. 1961, {Chapter-2 : Hydrodynamic and hydromagnetic
  stability} (International Series of Monographs on Physics, Oxford: Clarendon,
  1961)
%
\bibitem[\protect\citeauthoryear{{Chandrasekhar} \& {Kendall}}{{Chandrasekhar}
  \& {Kendall}}{1957}]{chandra57}
{Chandrasekhar}, S.,  \& {Kendall}, P.~C. 1957, \apj, 126, 457
%
\bibitem[\protect\citeauthoryear{{Chitre}}{{Chitre}}{1963}]{chit63}
{Chitre}, S.~M. 1963, \mnras, 126, 431
%
\bibitem[\protect\citeauthoryear{{Cuperman}, {Li}, \& {Semel}}{{Cuperman}
  et~al.}{1992}]{cupe92}
{Cuperman}, S., {Li}, J.,  \& {Semel}, M. 1992, \aap, 265, 296
%
\bibitem[\protect\citeauthoryear{{De Rosa} et~al.}{{De Rosa}
  et~al.}{2009}]{derosa09}
{De Rosa}, M.~L., {Schrijver}, C.~J., {Barnes}, G., et~al. 2009, \apj, 696, 1780
%
\bibitem[\protect\citeauthoryear{{Gary}}{{Gary}}{1989}]{gary89}
{Gary}, G.~A. 1989, \apjs, 69, 323
%
\bibitem[\protect\citeauthoryear{{Gary}}{{Gary}}{2001}]{gary01}
{Gary}, G.~A. 2001, \solphys, 203, 71
%
\bibitem[\protect\citeauthoryear{{Georgoulis} \& {LaBonte}}{{Georgoulis} \&
  {LaBonte}}{2004}]{geor04}
{Georgoulis}, M.~K.,  \& {LaBonte}, B.~J. 2004, \apj, 615, 1029
%
\bibitem[\protect\citeauthoryear{{Gosain}, {Tiwari}, \&
  {Venkatakrishnan}}{{Gosain} et~al.}{2010}]{gosain10}
{Gosain}, S., {Tiwari}, S.~K.,  \& {Venkatakrishnan}, P. 2010, \apj, 720, 1281
%
\bibitem[\protect\citeauthoryear{{Gosain}, {Venkatakrishnan}, \&
  {Tiwari}}{{Gosain} et~al.}{2009}]{gosain09}
{Gosain}, S., {Venkatakrishnan}, P.,  \& {Tiwari}, S.~K. 2009, \apjl, 706, L240
%
\bibitem[\protect\citeauthoryear{{Hagino} \& {Sakurai}}{{Hagino} \&
  {Sakurai}}{2004}]{hagi04}
{Hagino}, M.,  \& {Sakurai}, T. 2004, \pasj, 56, 831
%
\bibitem[\protect\citeauthoryear{{Harvey}}{{Harvey}}{1969}]{harv69}
{Harvey}, J.~W. 1969, Ph.D. thesis, University of Colorado, Boulder
%
\bibitem[\protect\citeauthoryear{{Ichimoto} et~al.}{{Ichimoto}
  et~al.}{2008}]{ichi08}
{Ichimoto}, K., {Lites}, B., {Elmore}, D., et~al. 2008, \solphys, 249, 233
%
\bibitem[\protect\citeauthoryear{{Kosugi} et~al.}{{Kosugi}
  et~al.}{2007}]{kosu07}
{Kosugi}, T., {Matsuzaki}, K., {Sakao}, T., et~al. 2007, \solphys, 243, 3
%
\bibitem[\protect\citeauthoryear{{Landolfi} \& {Landi
  Degl'Innocenti}}{{Landolfi} \& {Landi Degl'Innocenti}}{1982}]{lando82}
{Landolfi}, M.,  \& {Landi Degl'Innocenti}, E. 1982, \solphys, 78, 355
%
\bibitem[\protect\citeauthoryear{{Leka} \& {Skumanich}}{{Leka} \&
  {Skumanich}}{1999}]{leka99}
{Leka}, K.~D.,  \& {Skumanich}, A. 1999, \solphys, 188, 3
%
\bibitem[\protect\citeauthoryear{{Low}}{{Low}}{1973}]{low73a}
{Low}, B.~C. 1973, \apj, 184, 917
%
\bibitem[\protect\citeauthoryear{{Low}}{{Low}}{1974}]{low74a}
{Low}, B.~C. 1974, \apj, 193, 243
%
\bibitem[\protect\citeauthoryear{{Low}}{{Low}}{1977}]{low77a}
{Low}, B.~C. 1977, \apj, 217, 988
%
\bibitem[\protect\citeauthoryear{{Low}}{{Low}}{1978}]{low78}
{Low}, B.~C. 1978, \apj, 224, 668
%
\bibitem[\protect\citeauthoryear{{Low}}{{Low}}{1980}]{low80}
{Low}, B.~C. 1980, \apj, 239, 377
%
\bibitem[\protect\citeauthoryear{{Low}}{{Low}}{1982}]{low82a}
{Low}, B.~C. 1982, RvGSP, 20, 145
%
\bibitem[\protect\citeauthoryear{{Low}}{{Low}}{1985}]{low85}
{Low}, B.~C. 1985, In NASA. Marshall Space Flight Center Meas. of Solar Vector Magnetic Fields, 2374, 49
%
\bibitem[\protect\citeauthoryear{{Low} \& {Flyer}}{{Low} \&
  {Flyer}}{2007}]{low07}
{Low}, B.~C.,  \& {Flyer}, N. 2007, \apj, 668, 557
%
\bibitem[\protect\citeauthoryear{{Low} \& {Nakagawa}}{{Low} \&
  {Nakagawa}}{1975}]{low75}
{Low}, B.~C.,  \& {Nakagawa}, Y. 1975, in BAAS, Vol.~7, 347
%
\bibitem[\protect\citeauthoryear{{L{\"u}st} \& {Schl{\"u}ter}}{{L{\"u}st} \&
  {Schl{\"u}ter}}{1954}]{lust54}
{L{\"u}st}, R.,  \& {Schl{\"u}ter}, A. 1954, ZA, 34,
  263
%
\bibitem[\protect\citeauthoryear{{Mackay} \& {van Ballegooijen}}{{Mackay} \&
  {van Ballegooijen}}{2006a}]{mack06b}
{Mackay}, D.~H.,  \& {van Ballegooijen}, A.~A. 2006a, \apj, 641, 577
%
\bibitem[\protect\citeauthoryear{{Mackay} \& {van Ballegooijen}}{{Mackay} \&
  {van Ballegooijen}}{2006b}]{mack06a}
{Mackay}, D.~H.,  \& {van Ballegooijen}, A.~A. 2006b, \apj, 642, 1193
%
\bibitem[\protect\citeauthoryear{{Mackay} \& {van Ballegooijen}}{{Mackay} \&
  {van Ballegooijen}}{2009}]{mack09}
{Mackay}, D.~H.,  \& {van Ballegooijen}, A.~A. 2009, \solphys, 260, 321
%
\bibitem[\protect\citeauthoryear{{Mathew} et~al.}{{Mathew}
  et~al.}{2004}]{shibu04}
{Mathew}, S. K., {Solanki}, S. K., {Lagg}, A., et~al. 2004, \aap, 422, 693
%
\bibitem[\protect\citeauthoryear{{McClymont} \& {Mikic}}{{McClymont} \&
  {Mikic}}{1994}]{mccl94}
{McClymont}, A.~N.,  \& {Mikic}, Z. 1994, \apj, 422, 899
%
\bibitem[\protect\citeauthoryear{{Metcalf} et~al.}{{Metcalf}
  et~al.}{2008}]{metc08}
{Metcalf}, T.~R., {De Rosa}, M. L., {Schrijver}, C. J., et~al. 2008, \solphys, 247, 269
%
\bibitem[\protect\citeauthoryear{{Metcalf} et~al.}{{Metcalf}
  et~al.}{1995}]{metc95}
{Metcalf}, T.~R., {Jiao}, L., {McClymont}, A.~N., {Canfield}, R.~C.,  \&
  {Uitenbroek}, H. 1995, \apj, 439, 474
%
\bibitem[\protect\citeauthoryear{{Meyer}, {Schmidt}, \& {Weiss}}{{Meyer}
  et~al.}{1977}]{meye77}
{Meyer}, F., {Schmidt}, H.~U.,  \& {Weiss}, N.~O. 1977, \mnras, 179, 741
%
\bibitem[\protect\citeauthoryear{{Molodenskii}}{{Molodenskii}}{1969}]{molo69}
{Molodenskii}, M.~M. 1969, SovAe, 12, 585
%
\bibitem[\protect\citeauthoryear{{Molodensky}}{{Molodensky}}{1974}]{molo74}
{Molodensky}, M.~M. 1974, \solphys, 39, 393
%
\bibitem[\protect\citeauthoryear{{Moon} et~al.}{{Moon} et~al.}{2002}]{moon02}
{Moon}, Y., {Choe}, G.~S., {Yun}, H.~S., {Park}, Y.~D.,  \& {Mickey}, D.~L.
  2002, \apj, 568, 422
%
\bibitem[\protect\citeauthoryear{{Nakagawa} \& {Raadu}}{{Nakagawa} \&
  {Raadu}}{1972}]{naka72}
{Nakagawa}, Y.,  \& {Raadu}, M.~A. 1972, \solphys, 25, 127
%
\bibitem[\protect\citeauthoryear{{Parker}}{{Parker}}{1979}]{parker79}
{Parker}, E.~N. 1979, {Cosmical magnetic fields: Their origin and their
  activity} (Oxford, Clarendon Press; New York, Oxford University Press, 1979)
%
\bibitem[\protect\citeauthoryear{{Parker}}{{Parker}}{1989}]{parker89}
{Parker}, E.~N. 1989, GApFD, 45, 159
%
\bibitem[\protect\citeauthoryear{{Parker}}{{Parker}}{1990}]{parker90}
{Parker}, E.~N. 1990, GApFD, 52, 183
%
\bibitem[\protect\citeauthoryear{{Pevtsov}, {Canfield}, \& {Metcalf}}{{Pevtsov}
  et~al.}{1994}]{pcm94}
{Pevtsov}, A.~A., {Canfield}, R.~C.,  \& {Metcalf}, T.~R. 1994, \apjl, 425,
  L117
%
\bibitem[\protect\citeauthoryear{{Pevtsov}, {Canfield}, \& {Metcalf}}{{Pevtsov}
  et~al.}{1995}]{pcm95}
{Pevtsov}, A.~A., {Canfield}, R.~C.,  \& {Metcalf}, T.~R. 1995, \apjl, 440,
  L109
%
\bibitem[\protect\citeauthoryear{{Puschmann}, {Ruiz Cobo}, \& {Mart{\'{\i}}nez
  Pillet}}{{Puschmann} et~al.}{2010}]{pusc10}
{Puschmann}, K.~G., {Ruiz Cobo}, B.,  \& {Mart{\'{\i}}nez Pillet}, V. 2010,
  \apjl, 721, L58
%
\bibitem[\protect\citeauthoryear{{Rachkowsky}}{{Rachkowsky}}{1967}]{rach67}
{Rachkowsky}, D.~N. 1967, Izv. Krymsk. Astrofiz. Obs., 37, 56
%
\bibitem[\protect\citeauthoryear{{R{\'e}gnier} \& {Priest}}{{R{\'e}gnier} \&
  {Priest}}{2007}]{regn07}
{R{\'e}gnier}, S.,  \& {Priest}, E.~R. 2007, \aap, 468, 701
%
\bibitem[\protect\citeauthoryear{{Sakurai}}{{Sakurai}}{1979}]{saku79}
{Sakurai}, T. 1979, \pasj, 31, 209
%
\bibitem[\protect\citeauthoryear{{Sakurai}}{{Sakurai}}{1982}]{saku82}
{Sakurai}, T. 1982, \solphys, 76, 301
%
\bibitem[\protect\citeauthoryear{{Sakurai}}{{Sakurai}}{1989}]{saku89}
{Sakurai}, T. 1989, SSRv, 51, 11
%
\bibitem[\protect\citeauthoryear{{Sakurai}, {Makita}, \& {Shibasaki}}{{Sakurai}
  et~al.}{1985}]{saku85}
{Sakurai}, T., {Makita}, M.,  \& {Shibasaki}, K. 1985, MPA Rep., No.~212,
  p.~312 - 315
%
\bibitem[\protect\citeauthoryear{{Schmidt}}{{Schmidt}}{1964}]{schm64}
{Schmidt}, H.~U. 1964, NASA Special Publication, 50, 107
%
\bibitem[\protect\citeauthoryear{{Schrijver} et~al.}{{Schrijver}
  et~al.}{2008}]{schr08}
{Schrijver}, C.~J., {De Rosa}, M. L., {Metcalf}, T. R., et~al. 2008, \apj, 675, 1637
%
\bibitem[\protect\citeauthoryear{{Schrijver} et~al.}{{Schrijver}
  et~al.}{2006}]{schr06}
{Schrijver}, C.~J., {De Rosa}, M. L., {Metcalf}, T. R., et~al. 2006, \solphys, 235, 161
%
\bibitem[\protect\citeauthoryear{{Semel}}{{Semel}}{1967}]{semel67}
{Semel}, M. 1967, AnAp., 30, 513
%
\bibitem[\protect\citeauthoryear{{Shimizu} et~al.}{{Shimizu}
  et~al.}{2008}]{shim08}
{Shimizu}, T., {Nagata}, S., {Tsuneta}, S., et~al. 2008, \solphys, 249, 221
%
\bibitem[\protect\citeauthoryear{{Skumanich} \& {Lites}}{{Skumanich} \&
  {Lites}}{1987}]{skum87}
{Skumanich}, A.,  \& {Lites}, B.~W. 1987, \apj, 322, 473
%
\bibitem[\protect\citeauthoryear{{Solanki}, {Walther}, \&
  {Livingston}}{{Solanki} et~al.}{1993}]{sola93a}
{Solanki}, S.~K., {Walther}, U.,  \& {Livingston}, W. 1993, \aap, 277, 639
%
\bibitem[\protect\citeauthoryear{{Su} et~al.}{{Su} et~al.}{2009}]{su09}
{Su}, J.~T., {Sakurai}, T., {Suematsu}, Y., {Hagino}, M.,  \& {Liu}, Y. 2009,
  \apjl, 697, L103
%
\bibitem[\protect\citeauthoryear{{Suematsu} et~al.}{{Suematsu}
  et~al.}{2008}]{suem08}
{Suematsu}, Y., {Tsuneta}, S., {Ichimoto}, K., et~al. 2008, \solphys, 249, 197
%
\bibitem[\protect\citeauthoryear{{Tiwari}}{{Tiwari}}{2009}]{tiw09e}
{Tiwari}, S.~K. 2009, Ph.D. thesis, Udaipur Solar Observatory/Physical Research
  Laboratory, Mohanlal Sukhadia University, Udaipur
%
\bibitem[\protect\citeauthoryear{{Tiwari}}{{Tiwari}}{2010}]{tiw10}
{Tiwari}, S.~K. 2010, ArXiv:1009.5164
%
\bibitem[\protect\citeauthoryear{{Tiwari}, {Venkatakrishnan}, \&
  {Gosain}}{{Tiwari} et~al.}{2010}]{tiw10a}
{Tiwari}, S.~K., {Venkatakrishnan}, P.,  \& {Gosain}, S. 2010, \apj, 721, 622
%
\bibitem[\protect\citeauthoryear{{Tiwari} et~al.}{{Tiwari}
  et~al.}{2009a}]{tiw09a}
{Tiwari}, S.~K., {Venkatakrishnan}, P., {Gosain}, S.,  \& {Joshi}, J. 2009a,
  \apj, 700, 199
%
\bibitem[\protect\citeauthoryear{{Tiwari}, {Venkatakrishnan}, \&
  {Sankarasubramanian}}{{Tiwari} et~al.}{2009b}]{tiw09b}
{Tiwari}, S.~K., {Venkatakrishnan}, P.,  \& {Sankarasubramanian}, K. 2009b,
  \apjl, 702, L133
%
\bibitem[\protect\citeauthoryear{{Tsuneta} et~al.}{{Tsuneta}
  et~al.}{2008}]{tsun08}
{Tsuneta}, S., {Ichimoto}, K., {Katsukawa}, Y., et~al. 2008, \solphys, 249, 167
%
\bibitem[\protect\citeauthoryear{{Unno}}{{Unno}}{1956}]{unno56}
{Unno}, W. 1956, \pasj, 8, 108
%
\bibitem[\protect\citeauthoryear{{van Ballegooijen} \& {Cranmer}}{{van
  Ballegooijen} \& {Cranmer}}{2010}]{van10}
{van Ballegooijen}, A.~A.,  \& {Cranmer}, S.~R. 2010, \apj, 711, 164
%
\bibitem[\protect\citeauthoryear{{Venkatakrishnan}}{{Venkatakrishnan}}{1990a}]{venk90a}
{Venkatakrishnan}, P. 1990a, in IAU Symposium, Vol. 142, Basic Plasma Processes
  on the Sun, ed. {E.~R.~Priest \& V.~Krishan}, 323
%
\bibitem[\protect\citeauthoryear{{Venkatakrishnan}}{{Venkatakrishnan}}{1990b}]{venk90}
{Venkatakrishnan}, P. 1990b, \solphys, 128, 371
%
\bibitem[\protect\citeauthoryear{{Venkatakrishnan} \& {Gary}}{{Venkatakrishnan}
  \& {Gary}}{1989}]{venk89}
{Venkatakrishnan}, P.,  \& {Gary}, G.~A. 1989, \solphys, 120, 235
%
\bibitem[\protect\citeauthoryear{{Venkatakrishnan}, {Narayanan}, \&
  {Prasad}}{{Venkatakrishnan} et~al.}{1993}]{venk93}
{Venkatakrishnan}, P., {Narayanan}, R.~S.,  \& {Prasad}, N.~D.~N. 1993,
  \solphys, 144, 315
%
\bibitem[\protect\citeauthoryear{{Venkatakrishnan} \&
  {Tiwari}}{{Venkatakrishnan} \& {Tiwari}}{2009}]{venk09}
{Venkatakrishnan}, P.,  \& {Tiwari}, S.~K. 2009, \apjl, 706, L114
%
\bibitem[\protect\citeauthoryear{{Venkatakrishnan} \&
  {Tiwari}}{{Venkatakrishnan} \& {Tiwari}}{2010}]{venk10}
{Venkatakrishnan}, P.,  \& {Tiwari}, S.~K. 2010, \aap, 516, L5
%
\bibitem[\protect\citeauthoryear{{Wheatland} \& {R{\'e}gnier}}{{Wheatland} \&
  {R{\'e}gnier}}{2009}]{wheat09}
{Wheatland}, M.~S.,  \& {R{\'e}gnier}, S. 2009, \apjl, 700, L88
%
\bibitem[\protect\citeauthoryear{{Wiegelmann}}{{Wiegelmann}}{2004}]{wieg04}
{Wiegelmann}, T. 2004, \solphys, 219, 87
%
\bibitem[\protect\citeauthoryear{{Wiegelmann}, {Inhester}, \&
  {Sakurai}}{{Wiegelmann} et~al.}{2006}]{wieg06}
{Wiegelmann}, T., {Inhester}, B.,  \& {Sakurai}, T. 2006, \solphys, 233, 215
%
\bibitem[\protect\citeauthoryear{{Wiegelmann} et~al.}{{Wiegelmann}
  et~al.}{2005}]{wieg05}
{Wiegelmann}, T., {Lagg}, A., {Solanki}, S.~K., {Inhester}, B.,  \& {Woch}, J.
  2005, \aap, 433, 701
%
\bibitem[\protect\citeauthoryear{{Wiegelmann} et~al.}{{Wiegelmann}
  et~al.}{2010}]{wieg10}
{Wiegelmann}, T., {Yelles Chaouche}, L., {Solanki}, S.~K.,  \& {Lagg}, A. 2010,
  \aap, 511, A4
%
\bibitem[\protect\citeauthoryear{{Woltjer}}{{Woltjer}}{1958}]{wolt58}
{Woltjer}, L. 1958, \apj, 128, 384
%
\end{thebibliography}


\end{document}